\documentclass[aps,prx,twocolumn,showpacs,superscriptaddress,longbibliography]{revtex4-2}
\usepackage{amsfonts}
\usepackage{amsmath}
\usepackage{graphicx}
\usepackage{bm}
\usepackage{soul}
\usepackage{amssymb}
\usepackage{dcolumn}
\usepackage{color}
\usepackage{multirow}
\usepackage{booktabs}
\usepackage{subfigure}
\usepackage{array}
\usepackage{makecell}
\DeclareUnicodeCharacter{2212}{-}
\usepackage[colorlinks,
linkcolor=blue,
anchorcolor=blue,
citecolor=blue,
urlcolor=blue,
]{hyperref}

\begin{document}

    \title{Spin PN Junctions: Giant Magnetoresistance, Tunable Circular Polarization, and Spin Zener Filter}

    \author{Chun-Yi Xue}
    \affiliation{Kavli Institute for Theoretical Sciences, University of Chinese Academy of Sciences, Beijng 100049, China}

    \author{Gang Su}
    \email{gsu@ucas.ac.cn}
    \affiliation{Kavli Institute for Theoretical Sciences, University of Chinese Academy of Sciences, Beijng 100049, China}
    \affiliation{Physical Science Laboratory, Huairou National Comprehensive Science Center, Beijing 101400, China}
    \affiliation{Institute of Theoretical Physics, Chinese Academy of Sciences, Beijing 100190, China}
    \affiliation{School of Physical Sciences, University of Chinese Academy of Sciences, Beijng 100049, China}

    \author{Bo Gu}
    \email{gubo@ucas.ac.cn}
    \affiliation{Kavli Institute for Theoretical Sciences, University of Chinese Academy of Sciences, Beijng 100049, China}
    \affiliation{Physical Science Laboratory, Huairou National Comprehensive Science Center, Beijing 101400, China}

    \begin{abstract}
        We demonstrate that spin PN junctions—magnetic semiconductor homojunctions with spin-splitting-induced band offsets—fundamentally redefine carrier transport via spin-dependent recombination probabilities ($r_{\sigma\sigma^\prime}$). By integrating this mechanism into the Shockley model, we predict a 100× enhancement in magnetoresistance sensitivity under small forward bias, where exponential modulation of recombination lifetimes ($\tau_{\sigma\sigma^\prime}$) by magnetic fields amplifies resistance changes. Angular momentum conservation enables magnetically tunable circularly polarized luminescence: exclusive conduction-band or valence-band splitting in both neutral regions achieves near-half polarization ($\vert P_I\vert\approx50\%$), while global splitting degrades emission coherence. Furthermore, we propose a “spin Zener filter" exploiting $\approx1 eV$ valence band splitting in (Ga, Mn)As, where spin-dependent barrier heights generate near 100\% spin-polarized tunneling currents within a voltage-selective window. These results establish spin PN junctions as a universal design paradigm for magnetically amplified electronics, polarization-programmable optoelectronics, and voltage-gated spin injection without ferromagnetic contacts.

    \end{abstract}
    \pacs{}
    \maketitle


    \section{INTRODUCTION}
    Spintronics leverages the electron's spin degree of freedom to enable beyond-charge-based information technologies~\cite{RevModPhys.76.323,article,science,2}, yet faces a fundamental challenge: achieving efficient spin control at semiconductor interfaces. Magnetic semiconductors, exemplified by \mbox{(Ga, Mn)As}~\cite{PhysRevB.57.R2037,MEI2023101251,10.1063/1.2362971,PhysRevB.72.165204}, address this by integrating localized magnetic moments (via transition-metal doping) with semiconducting properties. Recent advances in Curie temperatures ($T_C$) and N-type magnetic semiconductors~\cite{mag1,mag2,mag3,mag4,mag5,cpl_40_6_067502} have expanded their technological viability, establishing these materials as critical platforms for spin manipulation.

    The generation, coherent transport, and relaxation of spin-polarized carriers form the foundation of spintronic device functionality~\cite{10.1063/1.102730,3,RevModPhys.76.323,4}.
    While enabling low-power logic and dense memory~\cite{Wolf2001Spintronics,2014The,8739466,spinDevice,Spingate}, key challenges persist: efficient spin injection, long-range diffusion stability, and room temperature compatibility. Magnetic semiconductors address these through exchange-mediated polarization~\cite{science,PhysRevB.57.R2037,MEI2023101251,10.1063/1.2362971,PhysRevB.72.165204}, with spin lifetime governed by spin-orbit coupling (SOC) and lattice symmetry (e.g., the Zener model~\cite{science}  and Dyakonov-Perel relaxation mechanism~\cite{3}). Recent advances in interfacial design~\cite{5,6} and topological hybrids enhance spin-current efficiency~\cite{topoPN0,topoPN,topo1,topo2}. Spin detection provides critical characterization by measuring the electron spin state (e.g., up/down) and its dynamics, through which physical mechanisms such as spin transport and relaxation can be revealed~\cite{electricalspin,SpinExp1,SpinExp2}. Crucially, angular-momentum-conserving recombination generates circularly polarized electroluminescence~\cite{Holub_2007,LED1,LED2,LED3}, enabling spin light-emitting diodes (spin-LEDs)—yet their performance remains bottlenecked by spin transport inefficiencies and material constraints~\cite{RevModPhys.76.323,7} at ambient conditions. 

    The PN junction—semiconductor technology's central element—remains underexploited for spintronic functionality. While spin-polarized PN junctions~\cite{PhysRevB.64.121201}, magnetic PN junctions~\cite{PhysRevLett.88.066603,PhysRevB.66.165301,PhysRevLett.97.026602}, and graphene PN junctions~\cite{grePN,SOCgphenePN} models exist, they universally don't assume spin-dependent recombination. In this work, we continue to employ the spin-polarized drift-diffusion model to further investigate the theory of spin-polarized carrier transport in magnetic PN junctions, which we refer to as spin PN junctions. The primary distinction between our work and previous studies lies in our hypothesis that the recombination probability in the transport equations is spin-dependent. Both external magnetic fields and magnetic doping can significantly enhance spin splitting ~\cite{8,9}. 
    Our central hypothesis establishes that spin splitting induces spin-dependent recombination as the dominant transport mechanism in low-bias conditions (nondegenerate regime), enabling new control dimensions beyond conventional models. Previous studies on spin-polarized transport in PN junctions primarily emphasized spin injection at the boundaries. In contrast, this work investigates spin-polarized transport in PN junctions through the application of an external magnetic field or via the inherent properties of magnetic materials, without relying on spin injection~\cite{PhysRevB.64.121201,PhysRevB.66.165301,PhysRevLett.88.066603}.
    
    Although quantum tunneling enables critical diode functionalities - Zener breakdown for voltage regulation~\cite{10}, Esaki negative differential resistance~\cite{11} and spin Esaki diodes~\cite{SPINESAKI}—we extend this physics to spintronics via the spin Zener filter. This device exploits spin-dependent barrier heights in magnetic heterostructures to achieve 100\% spin-polarized tunneling currents within a voltage-selective window. Unlike conventional Zener diodes, spin Zener filter creates an electrically gated spin filter,which can be a potential unit for high-efficiency spintronic circuits.   

     The content organization of this paper is structured as follows: SectionII introduces the detailed framework of theproposed model, including the band diagram of the device, the underlying assumptions and the transport equations for the spin PN junction. Section III presents the computational results of the spin-polarized PN junction: (i) 100× enhanced magnetoresistance sensitivity, (ii) magnetically tunable near-half circular polarization, and (iii) voltage-controlled spin filtering via the spin Zener filter. Section IV, as the concluding part of this work, briefly summarizes the key theoretical contributions and computational findings of this study.
	
     \section{MODEL}
      \begin{figure*}[tphb!]
    	\centering
            \includegraphics[width=\textwidth]{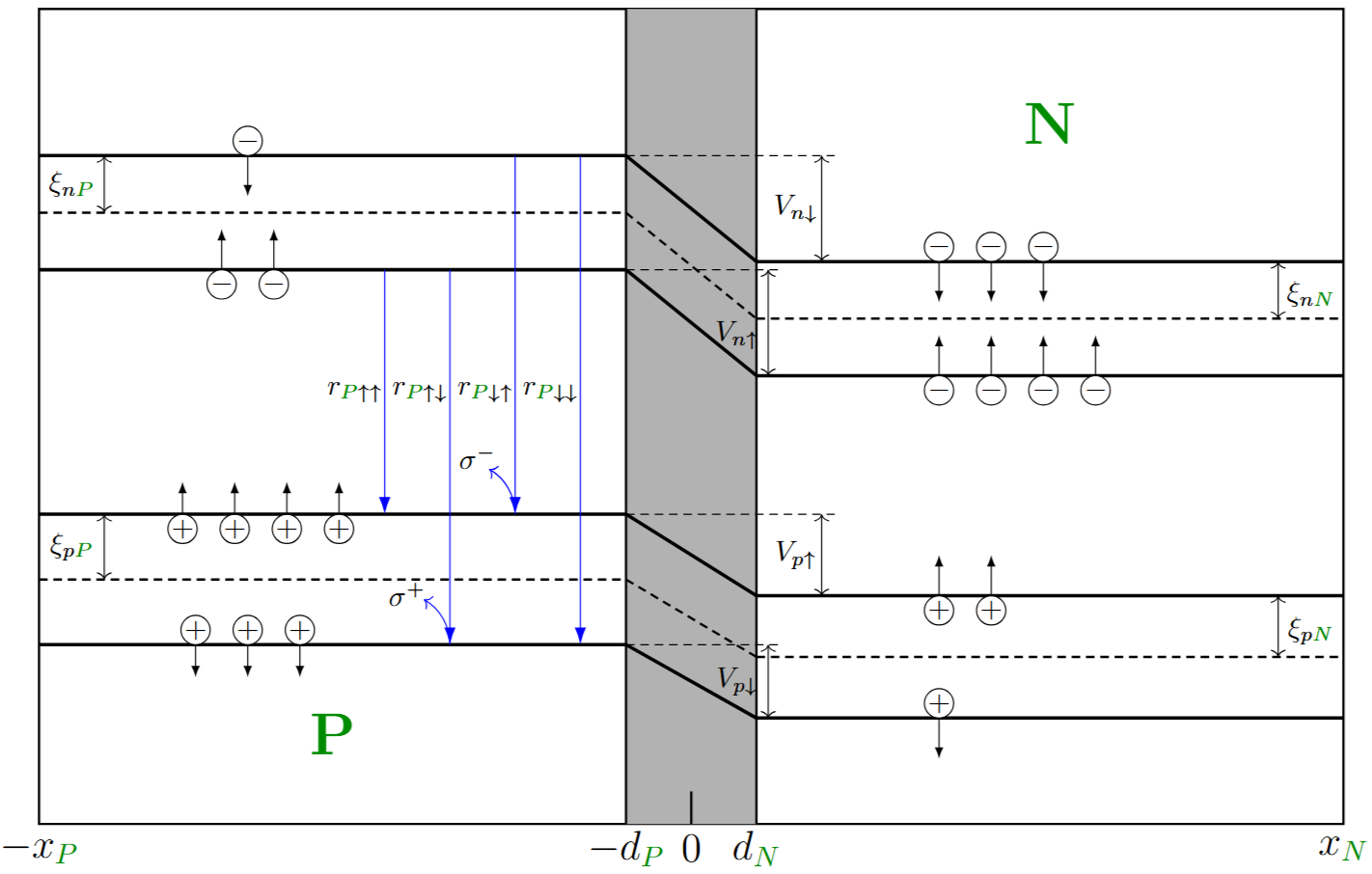}
    	\caption{
        Band scheme of Spin PN junction. In this PN junction, both the p-type and n-type regions are composed of magnetic materials. This junction is P doped from $-x_P$ to $0$ and N doped from 0 to $x_N$. The depletion layer forms at $-d_P$ to $d_N$. The conduction band and valence band spin splitting in the P region is $2\xi_{nP}$ and $2\xi_{pP}$. The green labels “P" and “N" denote the regions, while the black labels “n" and “p" represent electrons and holes, respectively. The intrinsic effective built-in field across the depletion layer is $V_0$. The spin-polarized carriers' built in fields become spin dependent: $V_{n\uparrow}=V_{0}+\xi_{\rm nN}-\xi_{\rm nP}$,$V_{n\downarrow}=V_{0}-\xi_{\rm nN}+\xi_{\rm nP}$,$V_{p\uparrow}=V_{0}+\xi_{\rm pP}-\xi_{\rm pN}$ and $V_{p\downarrow}=V_{0}-\xi_{\rm pP}+\xi_{\rm pN}$. The left side corresponds to the P neutral region, the gray area represents the depletion layer, and the right side is the N neutral region. The diagram also illustrates the distinct built-in potential barriers for carriers with different spin polarizations, as well as their corresponding spin splitting energies. Due to spin splitting, the energy levels of carriers with different spins are distinct. This results in varying recombination probabilities when spin-polarized electrons and holes undergo recombination. As illustrated in the figure, we will introduce four spin-dependent recombination probabilities. Conserving angular momentum and neglecting orbital contributions, the recombination of differently spin-polarized electron-hole pairs generates circularly polarized light.}
    	\label{fig_band}
    \end{figure*}
     \begin{table}[htbp]
    \centering
    \caption{Summary of the notation used in the text.The equilibrium notation are shown in brackets}
    \label{notation}
        \begin{tabular}{cl}  
        \hline\hline                 
        $r_{\sigma\sigma^\prime}$ & \makecell[l]{recombination probability between $\sigma$ electron and \\$\sigma^\prime$ hole}\\ \hline
        $r_{P(N)\sigma\sigma^\prime}$ & \makecell[l]{recombination probability between $\sigma$ electron and\\ $\sigma^\prime$ hole in P(N) region}\\ \hline
        $r_0$ & \makecell[l]{recombination probability without spin splitting}\\ \hline
        $N_{D\sigma}$ & $\sigma$ density of donors in N region  \\ \hline 
        $N_{A\sigma^\prime}$ & $\sigma^\prime$ density of acceptors in P region  \\ \hline
         $N_{D}$ & total donor density in N region  \\ \hline 
        $N_{A}$ & total acceptor density in P region  \\ \hline 
        $\tau_{n\sigma\sigma^\prime}$ & \makecell[l]{electron lifetime for $\sigma$ electron in recombination \\ with $\sigma^\prime$ hole in P region ($\tau_{n\sigma\sigma^\prime}={1}/{r_{\sigma\sigma^\prime}N_{A\sigma^\prime}}$) }\\ \hline 
        $\tau_{p\sigma\sigma^\prime}$ & \makecell[l]{hole lifetime for $\sigma^\prime$ hole in recombination with\\ $\sigma$ electron in N region ($\tau_{p\sigma\sigma^\prime}={1}/{r_{\sigma\sigma^\prime}N_{D\sigma}}$ )}\\ \hline 
        $\xi_{n/p,N/P}$ & \makecell[l]{conduction(n) and valence band(p) splitting in N/P \\ region} \\ \hline
         $\xi_{n/p,N/P0}$ & \makecell[l]{intrinsic conduction(n) and valence band(p) splitting \\ in N/P region} \\ \hline
        $\delta\xi_{n/p,N/P}$ & \makecell[l]{conduction(n) and valence band(p) splitting induced\\ by external magnetic field N/P region}\\ \hline
        $n_\sigma$ & density of $\sigma$ electrons (equilibrium $n_{\sigma0}$)  \\ \hline 
        $n$ & total electron density ($n=\sum\limits_{\sigma}{n_\sigma}$, equilibrium $n_0$)  \\ \hline 
        $p_{\sigma^\prime}$ & density of $\sigma^\prime$  holes (equilibrium $p_{\sigma^\prime0}$)  \\ \hline 
        $p$ & total hole density ($p=\sum\limits_{\sigma^\prime}{p_{\sigma^\prime}}$, equilibrium $p_0$)\\ \hline 
        $\alpha_n$ & spin polarization of electrons($\alpha_{n0}$)\\ \hline 
        $\alpha_p$ & spin polarization of holes($\alpha_{p0}$)\\ \hline 
        $j_{n\sigma\sigma^\prime}$ & \makecell[l]{electron charge current due to recombination of $\sigma$ \\electron with $\sigma^\prime$ hole}\\ \hline 
        $j_{p\sigma\sigma^\prime}$ & \makecell[l]{hole charge current due to recombination of $\sigma$ ele-\\ctron and $\sigma^\prime$ hole}\\ \hline 
        $j_{n\sigma}$ & charge current of $\sigma$ spin electrons \\ \hline 
        $P_I$ & circular polarization of light \\ \hline
        $P_j$ & spin polarization of charge current  \\ \hline
        $J_n$ & electron particle current\\ \hline 
        $j_n$ & electron charge current\\ \hline 
        $j_p$ & hole charge current\\ \hline 
        $j$ & total charge current \\ \hline  
        $D_{nP\sigma }$ & $\sigma$ electron diffusivity in P region \\ \hline 
        $D_{pN\sigma^\prime }$ & $\sigma^\prime$ hole diffusivity in N region\\  \hline 
        $L_{nP\sigma\sigma^\prime}$ & \makecell[l]{electron diffusion length in a $\sigma$ electron $\sigma^\prime$ hole \\recombination process in P region \\($L_{nP\sigma\sigma^\prime}=\sqrt{D_{nP\sigma}\tau_{nP\sigma\sigma^\prime}}$)} \\ \hline 
        $L_{pN\sigma\sigma^\prime}$ & \makecell[l]{hole diffusion length in a $\sigma$ electron $\sigma^\prime$ hole reco-\\mbination process in N region. \\($L_{pN\sigma\sigma^\prime}=\sqrt{D_{pN\sigma^\prime}\tau_{pN\sigma\sigma^\prime}}$)}\\ 
         \hline\hline 
        \end{tabular}
    \end{table}
     \begin{table*}[htbp]
    \centering
    \caption{Difference among spin PN junction, magnetic PN junction and PN junction.}
    \begin{tabular}{c|ccc}
    \hline \hline
     Model & \makecell{Intrinsic band splittings} &\makecell{Band splittings induced\\ by external magnetic field } & \makecell{ Recombination probability }    \\  
     \hline \makecell{Spin PN junction} & \makecell{$\xi_{nP0}\neq0$,$\xi_{nN0}\neq0$\\$\xi_{pP0}\neq0$,$\xi_{pN0}\neq0$}  & \makecell{$\delta\xi_{nP}\neq0$,$\delta\xi_{nN}\neq0$\\$\delta\xi_{pP}\neq0$,$\delta\xi_{pN}\neq0$} & \makecell{$r_{\uparrow\uparrow} \neq r_{\uparrow\downarrow}\neq r_{\downarrow\uparrow} \neq r_{\downarrow\downarrow}$}  \\ 
     \hline \makecell{Magnetic PN junction‌~\cite{PhysRevLett.88.066603}} & \makecell{$\xi_{nP0}=0$,$\xi_{nN0}=0$\\$\xi_{pP0}=0$,$\xi_{pN0}=0$} & \makecell{$\delta\xi_{nP}\neq0$,$\delta\xi_{nN}=0$\\$\delta\xi_{pP}=0$,$\delta\xi_{pN}=0$} & $r_{\uparrow\uparrow}= r_{\uparrow\downarrow}=r_{\downarrow\uparrow} =r_{\downarrow\downarrow}$   \\ \hline \makecell{PN junction} &\makecell{$\xi_{nP0}=0$,$\xi_{nN0}=0$\\$\xi_{pP0}=0$,$\xi_{pN0}=0$} & \makecell{$\delta\xi_{nP}=0$,$\delta\xi_{nN}=0$\\$\delta\xi_{pP}=0$,$\delta\xi_{pN}=0$} &  $r_{\uparrow\uparrow}= r_{\uparrow\downarrow}=r_{\downarrow\uparrow} =r_{\downarrow\downarrow}$ 
      \\ \hline \hline
    \end{tabular}	
    \label{table2}
    \end{table*}
    ‌
   
   We model a symmetric PN junction with magnetic semiconductors as shown in Fig.~\ref{fig_band}, where spin splitting $\xi_{n/p,N/P}$ manifests in both conduction (n) and valence (p) bands of P and N regions. The spin splitting in materials can originate from both intrinsic splitting $\xi_{n/p,N/P0}$ in magnetic materials and Zeeman splitting induced by external magnetic fields $\delta\xi_{n/p,N/P}$. In general, the g-factor varies with different charge carriers (electrons, holes) and spatial positions under non-uniform doping conditions. However, in our calculations, for simplicity, the g-factor is treated as a constant. Zeeman splitting induced by external magnetic fields follow the Zeeman form $2\delta\xi_{n/p,N/P} = g\mu_BB$, with magnetic doping enhancing g-factors for amplified response. The total spin splitting is the sum of the intrinsic spin splitting energy and the energy induced by an external field 
   \begin{equation} 
   \xi_{n/p,N/P}=\xi_{n/p,N/P0}+\delta\xi_{n/p,N/P}.
    \end{equation}Magnetic doping can significantly enhance spin splitting.
    
    Crucially, we introduce spin-dependent recombination probabilities induced by spin-splitting. The spin-dependent recombination probability creates distinct recombination pathways for each spin configuration ($\uparrow\uparrow, \uparrow\downarrow, \downarrow\uparrow, \downarrow\downarrow)$, fundamentally differentiating our framework from spin-independent models~\cite{PhysRevB.66.165301,PhysRevLett.88.066603}. Orbital effects are neglected throughout, restricting dynamics to pure spin degrees of freedom.

    Within the established drift-diffusion framework for magnetic PN junctions~\cite{PhysRevB.66.165301,PhysRevLett.88.066603}, we model low forward-bias conditions where carrier transport balances drift, diffusion, recombination, and spin relaxation. Our fundamental advance incorporates spin-dependent recombination probabilities $r_{\sigma\sigma^\prime}$—unequal due to spin-splitting-induced energy differences—into the steady-state continuity equations. For neutral regions under contact injection, non-equilibrium carriers undergo simultaneous diffusion-driven recombination and spin relaxation, governed by

    \begin{equation} 
    J_{n\uparrow}^{\prime}+J_{n\downarrow}^{\prime}=-\sum\limits_{\sigma\sigma^\prime}r_{\sigma\sigma^\prime}(n_{\sigma}p_{\sigma^\prime}-n_{\sigma0}p_{\sigma^\prime0}),
    \end{equation}
    \begin{equation}
    \begin{aligned}
    & J_{n\uparrow}^{\prime}-J_{n\downarrow}^{\prime}=-\sum\limits_{\sigma^\prime}r_{\uparrow\sigma^\prime}(n_{\uparrow}p_{\sigma^\prime}-n_{\uparrow0}p_{\sigma^\prime0})+\\&\sum\limits_{\sigma^\prime}r_{\downarrow\sigma^\prime}(n_{\downarrow}p_{\sigma^\prime}-n_{\downarrow0}p_{\sigma^\prime0})-\frac{\delta n_\uparrow-\delta n_\downarrow-\alpha_{n0}(\delta n_\uparrow+\delta n_\downarrow)}{T_{1n}},
     \end{aligned}
     \end{equation}
    where $r_{\uparrow\uparrow}$,$r_{\uparrow\downarrow}$,$r_{\downarrow\uparrow}$,$r_{\downarrow\downarrow}$ is the spin dependent recombination rate, the first index corresponds to the electron spin, and the second index corresponds to the hole spin.
    The current of electrons can be written as $J_{n\uparrow}=-D_{n\uparrow}n_{\uparrow}^{\prime}$ and $J_{n\downarrow}=-D_{n\downarrow}n_{\downarrow}^{\prime}$.
    By combining the steady-state carrier diffusion equation with the spin relaxation equation (1) and(2), we derive the diffusion equation for spin-polarized electrons in the neutral region,
    \begin{equation}
    \begin{aligned}
    \delta n_{\uparrow}^{\prime\prime}=\frac{\delta n_{\uparrow}}{D_{n\uparrow}\tau_{n\uparrow\uparrow}}
    +\frac{\delta n_{\uparrow}}{D_{n\uparrow}\tau_{n\uparrow\downarrow}}
    +\frac{\delta n_{\uparrow}-\delta n_{\downarrow}-\alpha_{n0}(\delta n_\uparrow+\delta n_\downarrow)}{2D_{n\uparrow}T_{1n}},
    \end{aligned}
    \end{equation}
    \begin{equation}
    \begin{aligned}
    \delta n_{\downarrow}^{\prime\prime}=\frac{\delta n_{\downarrow}}{D_{n\downarrow}\tau_{n\downarrow\downarrow}}
    +\frac{\delta n_{\downarrow}}{D_{n\downarrow}\tau_{n\downarrow\uparrow}}
    -\frac{\delta n_{\uparrow}-\delta n_{\downarrow}-\alpha_{n0}(\delta n_\uparrow+\delta n_\downarrow)}{2D_{n\downarrow}T_{1n}},
    \end{aligned}
    \end{equation}
    where $\tau_{n\uparrow\uparrow}$,$\tau_{n\uparrow\downarrow}$,$\tau_{n\downarrow\uparrow}$,$\tau_{n\downarrow\downarrow}$ are the electron lifetimes for specific recombination processes, which are defined as $\tau_{n\sigma\sigma^\prime}={1}/{r_{\sigma\sigma^\prime}p_{\sigma^\prime}}$. The transport equations for spin-polarized holes are provided in the Supplemental Material~\cite{Supply}.
    Based on the thermal equilibrium between excitation and recombination in direct bandgap semiconductors, we assume that spin-splitting gives rise to distinct spin-dependent recombination probabilities
    \begin{equation}
    r_{\sigma\sigma^\prime}=r_0\exp[\frac{(E_{C\sigma}-E_{V{\sigma^\prime}})-(E_{C0}-E_{V0})}{k_BT}],
    \end{equation}
    where $r_0$ is the recombination probability in the absence of spin splitting, $[(E_{C\sigma}-E_{V{\sigma^\prime}})-(E_{C0}-E_{V0})]$ is the spin-splitting-induced band offset.
    
    In the treatment of the depletion layer, we assume that the electrostatic potential generated by the applied voltage is $\chi(x)$, and that the entire voltage drop across the PN junction is confined to the depletion layer. This allows us to establish the relationship between the electrostatic potentials at the left and right edges of the depletion layer due to the applied voltage
     \begin{equation}
    \chi(-d_N)-\chi(d_P)=V.
    \end{equation}
    For the P and N neutral regions, $\chi$ is constant. By adopting ‌Shockley's constant chemical potential approximation‌~\cite{15}, which assumes that the quasi-chemical potential remains spatially invariant within the depletion layer, we derive the relationship between the electron densities at the two edges of the depletion layer
    \begin{equation}
    \frac{n_{\uparrow}(-d_P)}{n_{\uparrow}(d_N)}=\exp(\frac{-e\xi_{nN}+e\xi_{nP}-eV_0+eV}{k_BT}),
    \end{equation}
    \begin{equation}
    \frac{n_{\downarrow}(-d_P)}{n_{\downarrow}(d_N)}=\exp(\frac{e\xi_{nN}-e\xi_{nP}-eV_0+eV}{k_BT}),
    \end{equation}
    where $V_0$ is built-in voltage in the absence of an external bias.  For detailed derivations of the equations of carrier densities in the depletion layer, please refer to Supplemental Material~\cite{Supply}.
    
     The current in a spin PN junction under forward bias and low-biases conditions, as discussed in this work, is primarily driven by the electric field. However, an external magnetic field also influences the current magnitude. This arises because the external magnetic field modifies the energy levels of charge carriers, thereby altering their densities and electron-hole recombination probabilities, as shown in Eq.(6). We neglect recombination currents within the depletion layer and focus solely on the recombination of non-equilibrium minority carriers in the neutral regions. For each recombination process, the electron recombination current $j_{n\sigma\sigma^\prime}$ can be written as 
     \begin{equation}
    j_{n\sigma\sigma^\prime}= e\frac{D_{nP\sigma}}{L_{nP\sigma\sigma^\prime}}n_{0P\sigma}\coth(\frac{x_P-d_P}{L_{nP\sigma\sigma^\prime}})[\exp(\frac{eV}{k_BT})-1],
    \end{equation}
    where $D_{nP\sigma}$ is the diffusivity of $\sigma$ electrons in P region; $L_{nP\sigma}$ is the diffusion length associated with recombination process between $\sigma$ electrons and $\sigma^\prime$ holes in P region, which is defined as $L_{nP\sigma\sigma^\prime}=\sqrt{D_{nP\sigma}\tau_{nP\sigma\sigma^\prime}}$, where the electron lifetime of a specific recombination process $\tau_{nP\sigma\sigma^\prime}$ is defined as $\tau_{nP\sigma\sigma^\prime}={1}/{r_{\sigma\sigma^\prime}N_{A\sigma^\prime}}$ .
     The spin polarized electron current $j_{n\sigma}$ can be obtained by summing over the hole indices
     \begin{equation}
    j_{n\sigma}=\sum\limits_{\sigma^\prime} e\frac{D_{nP\sigma}}{L_{nP\sigma\sigma^\prime}}n_{0P\sigma}\coth(\frac{x_P-d_P}{L_{nP\sigma\sigma^\prime}})[\exp(\frac{eV}{k_BT})-1].
    \end{equation}
     The total electron current $j_n$ is the sum of spin-up and spin-down currents
     \begin{equation}
         j_n=\sum\limits_{\sigma} j_{n\sigma}.
     \end{equation}
     Following the derivation for spin-polarized electrons, the hole current can be expressed as
     \begin{equation}
          j_{p\sigma^\prime}=\sum\limits_{\sigma} e\frac{D_{pN\sigma^\prime}}{L_{pN\sigma\sigma^\prime}}p_{0N\sigma^\prime}\coth(\frac{x_N-d_N}{L_{pN\sigma\sigma^\prime}})[\exp(\frac{eV}{k_BT})-1] ,
     \end{equation}
      \begin{equation}
         j_p=\sum\limits_{\sigma^\prime} j_{p\sigma^\prime},
     \end{equation}
      where $D_{pN\sigma^\prime}$ is the diffusivity of $\sigma^\prime$ hole in N region; $L_{pN\sigma\sigma^\prime}$ is the diffusion length of recombination process between $\sigma$ electrons and $\sigma^\prime$ holes in N region, which is defined as $L_{pN\sigma\sigma^\prime}=\sqrt{D_{pN\sigma^\prime}\tau_{pN\sigma\sigma^\prime}}$.
      
     The total current through the PN junction is the sum of the electron current and the hole current
     \begin{equation}
         j=j_n+j_p.
     \end{equation}
    External magnetic field can modify the magnitude of spin splitting, thereby altering the energy levels of spin-polarized conduction and valence bands. This in turn modulates carrier density and recombination probability, ultimately enabling precise regulation of electrical current. The magnetoresistance (MR) of spin PN junction is define as
    \begin{equation}
         MR=\frac{R(B)-R(B=0)}{R(B=0)}=\frac{\frac{V}{j(B)}-\frac{V}{j(B=0)}}{\frac{V}{j(B=0)}}.
     \end{equation}

    The model discussed in this work neglects orbital effects; therefore, the magnetic quantum numbers of the carriers originate exclusively from their spin quantum numbers. The spin-polarized luminescence process in the spin PN junction arises from electron-hole recombination. Spin-splitting results in distinct energy level differences between electrons and holes with different spin. When electrons and holes with opposite spin polarizations recombine, they emit circularly polarized light, as shown in Fig.~\ref{fig_band}. Consequently, the degree of circular polarization of the emitted light depends on the relative contributions of recombination processes that involve carriers with opposite spin
    \begin{equation}
    P_I=\frac{I(\sigma^+)-I(\sigma^-)}{I(\sigma^+)+I(\sigma^-)+I_{unpolarized}}=\frac{I(\sigma^+)-I(\sigma^-)}{I_{total}},
    \end{equation}
    where $I(\sigma^+)$ and $I(\sigma^-)$ represent the luminescence intensities of right-handed and left-handed circularly polarized light,$P_I$ is the circular polarization degree of light. Our model considers only the spin angular momentum of the carriers. According to the principle of angular momentum conservation, the recombination of a spin-up electron with a spin-down hole emits right-handed circularly polarized light, while conversely, the recombination of a spin-down electron with a spin-up hole emits left-handed circularly polarized light. Consequently, the expression for luminescence intensity can be represented by the following relationship
   \begin{equation}
   I(\sigma^+)=r_{P\uparrow\downarrow}n_{P\uparrow}N_{A\downarrow}+r_{N\uparrow\downarrow}N_{D\uparrow}p_{N\downarrow},
   \end{equation}
   \begin{equation}
    \begin{aligned}
   I(\sigma^-)=r_{P\downarrow\uparrow}n_{P\downarrow}N_{A\uparrow}+r_{N\downarrow\uparrow}N_{D\downarrow}p_{N\uparrow},
   \end{aligned}
   \end{equation}
   \begin{equation}
   \begin{aligned}
        I_{total}=\sum\limits_{\sigma\sigma^\prime}(r_{P\sigma\sigma^\prime}n_{P\sigma}N_{A\sigma^\prime}+r_{N\sigma\sigma^\prime}N_{D\sigma}p_{N\sigma^\prime}).
    \end{aligned}
   \end{equation}
   \begin{figure*}[htbp!]
    	\centering
            \subfigure{\includegraphics[width=0.455\textwidth]{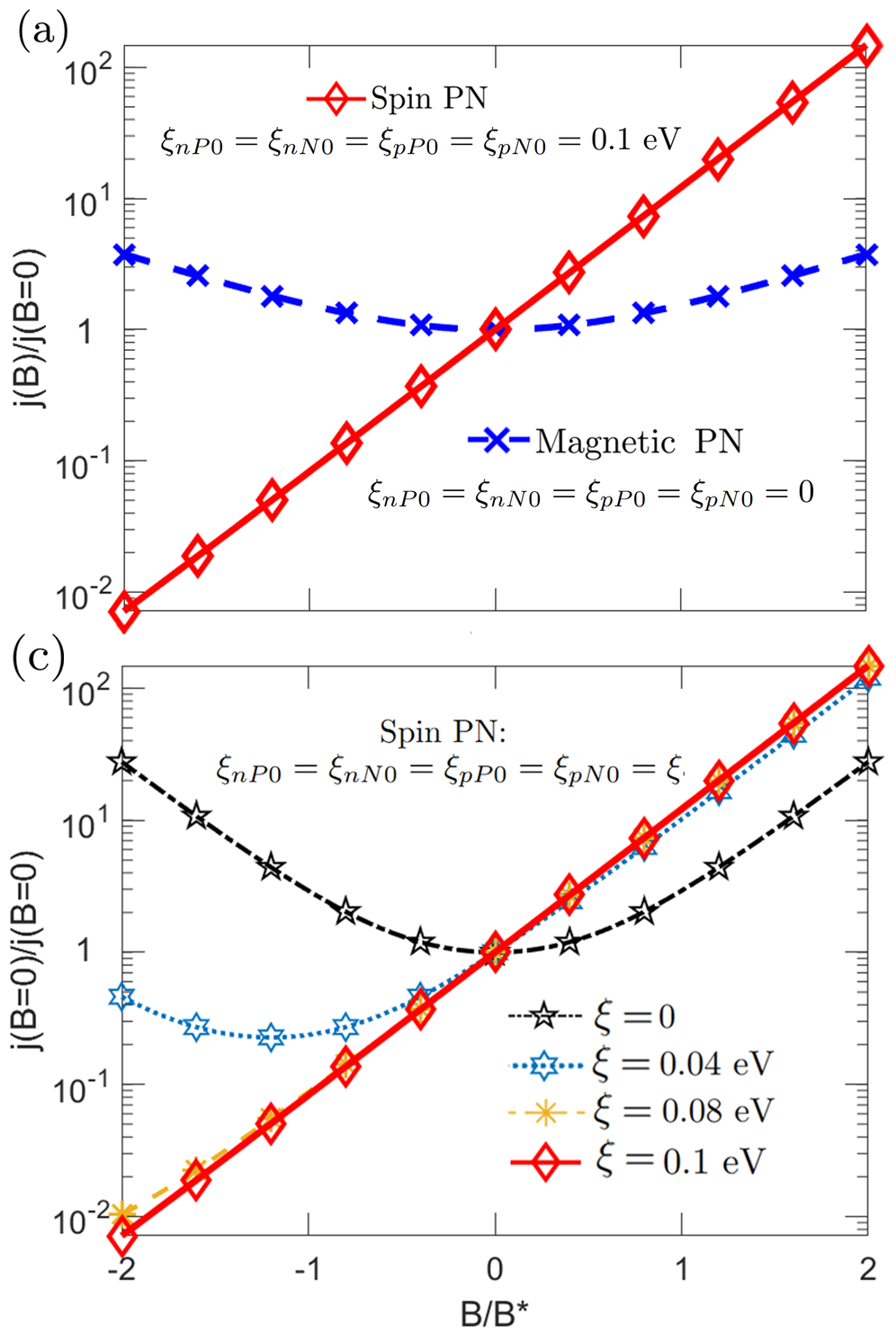}}
            \subfigure{\includegraphics[width=0.45\textwidth]{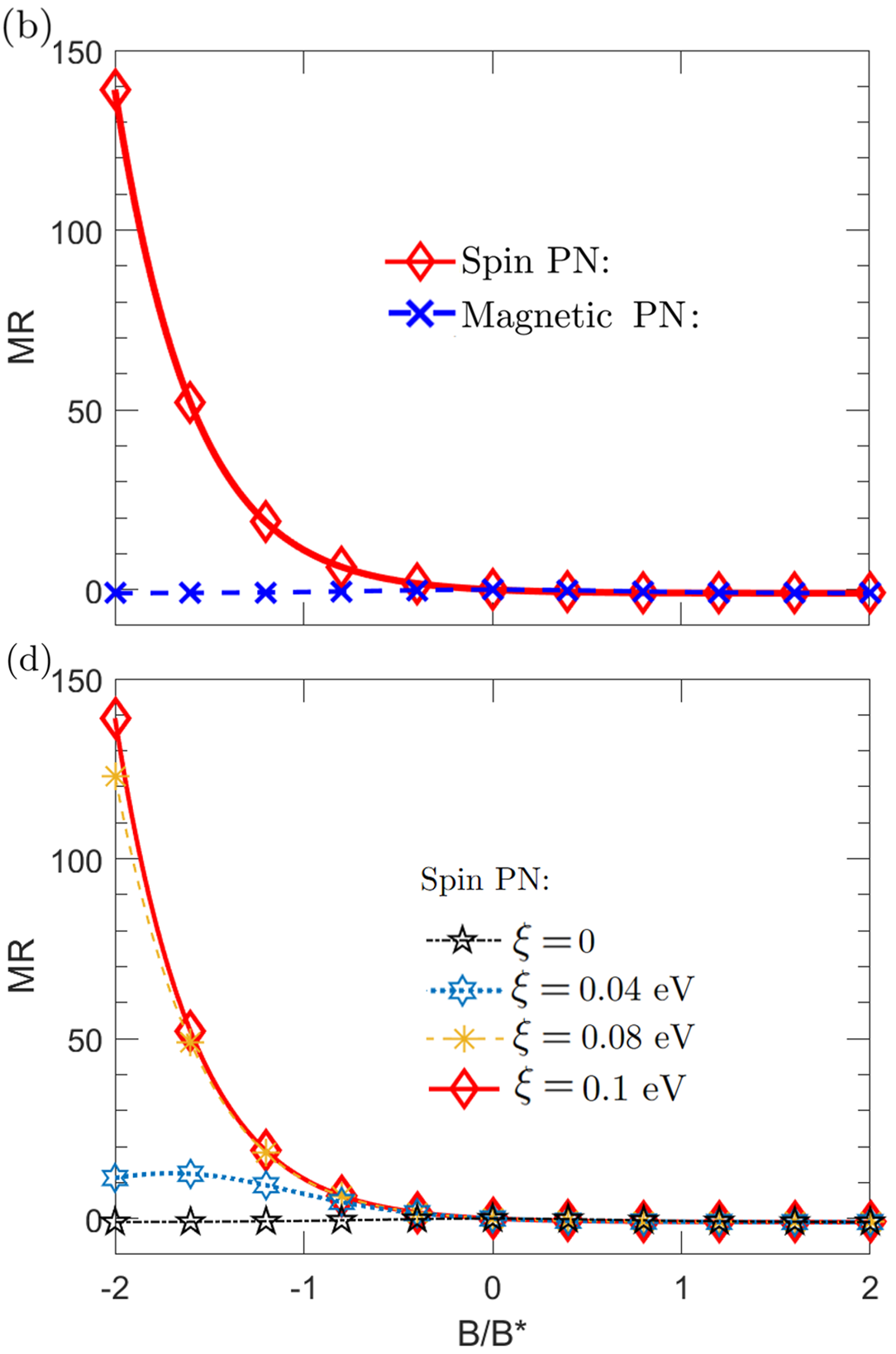}} \\
    	\caption{
        Giant magnetoresistance and current modulation in spin PN junctions. (a) Relative current magnitude $j(B)/j(0)$ as a function of external magnetic field B (scaled to $B^*={2k_BT}/{g\mu_B}$). The spin PN junction (red solid curve) exhibits exponentially enhanced sensitivity compared to conventional magnetic PN junctions (blue dashed curve) under small forward bias ($V=\rm{+0.5V}$), owing to spin-dependent recombination probabilities $r_{\sigma\sigma^\prime}$. (b) Magnetoresistance MR (defined by Eq. (16)) for spin PN junctions (red) and magnetic PN junctions (blue). The spin PN junction achieves $100\times$ enhanced MR sensitivity due to magnetic-field-tunable recombination lifetimes. (c) Current modulation sensitivity to intrinsic band splitting $\xi$ . For $\xi>0.1\rm{eV}$, current saturates as spin splitting dominates recombination dynamics. (d) MR modulation sensitivity to intrinsic band splitting $\xi$. The maximum MR response occurs at $\xi=0.1 \rm{eV(\approx4k_BT}$ at 300 K), beyond which incremental gains diminish. Parameters: Symmetric spin PN junction with $N_A=N_D=10^{16}\rm{cm^{-3}}, \xi_{n/p,N/P0}=0.1\rm{eV}, D_{nP}=100\rm{cm^2/s}, D_{pN}=\rm{10cm^2/s}, r_0=3\times10^{−5}\rm{cm^3/s}, T=300\rm{K}$.}
        \label{fig_current}
    \end{figure*}

    \section{RESULTS AND DISCUSSION}
    {
    We quantitatively benchmark our spin-PN junction model against prior magnetic PN junction theories~\cite{PhysRevB.66.165301,PhysRevLett.88.066603} and classical PN junctions, with key distinctions summarized in table.~\ref{table2}. Our framework uniquely incorporates spin-dependent recombination in magnetic semiconductors. Magnetic field strengths are scaled to

    \begin{equation}
        B^*=\frac{2k_BT}{g\mu_B}
    \end{equation}
    to enable material-agnostic analysis. Some materials exhibit large g-factors, such as InSb, where the g-factor can reach 30–50 at 1.5K~\cite{gf1}($B^*\approx 0.2\rm{T}$), and even up to 100 in its magnetic superlattice structures at 5 K~\cite{gf2}($B^*\approx 0.2\rm{T}$). For large g-factor materials, magnetic fields can significantly modulate spin splitting. However, at room temperature, its g-factor decreases significantly. It is noted that high Curie temperature was discussed for Fe doped InSb both experimentally and theoretically \cite{Tu_2019,PhysRevB.102.094432}.  
 
    }
    \subsection{Giant magnetoresistance and current modulation}
    
    {
    Our numerical simulations employ a symmetric magnetic PN junction with intrinsic band splitting $\xi_{nP0}=\xi_{nN0}=\xi_{pP0}=\xi_{pN0}=0.1\rm{ eV(\approx4k_BT}$ at 300K) in both regions. The calculation parameters selected are the key parameters of the magnetic PN junction~\cite{PhysRevB.66.165301,PhysRevLett.88.066603}: $N_A=N_D=10^{16}\rm{cm^{-3}}$,$\rm{n_i=1.8\times10^6cm^{-3}}$, spin-degenerate diffusivities $D_{nP\uparrow}=D_{nP\downarrow}=10D_{pN\uparrow}=10D_{pN\downarrow}=100\rm{cm^2/s}$, $r_0=3\times10^{-5}\rm{cm^3/s}$ ($r_{\sigma\sigma^\prime}$ defined by Eq.(6)), device size $x_{P}=x_{N}=3\mu m$, and forward bias $V=+0.5\rm{V}$. Using Eqs. (10)-(16), we calculate magnetic-field-dependent current and MR (Eq.(16)). Fig.~\ref{fig_current}(a), (b) demonstrates a 10-100× enhancement in both quantities versus magnetic PN junction models~\cite{PhysRevB.66.165301,PhysRevLett.88.066603}—directly attributable to spin-dependent recombination.  The results presented here demonstrate the magnetic field modulation behavior when the intrinsic band splittings reach 0.1 eV. This orders-of-magnitude amplification resolves the critical sensitivity limitations of existing spintronic devices, establishing spin-dependent recombination as a new control mechanism for magnetic-field-tunable electronics. The enhancement mechanism arises from B-field modulation of band splitting, exponentially increasing recombination rates for specific spin configurations. As shown in Fig.~\ref{fig_current}(c), (d) for the parameter sensitivity of current and MR modulation by magnetic field to band splitting in the absence of an external field, it can be seen that ‌when the intrinsic spin splitting of the material itself exceeds 0.1 eV, the relative change in current no longer increases significantly with magnetic field modulation‌.

    The magnetic field modulation effect is conserved in both the relative current and MR, even for high doping concentrations like $N_A =N_D =10^{19} \rm{cm^{−3}}$.
    }
    \subsection{Magnetically tunable circular polarization}
    {
    We compute the circular polarization degree $P_I$ (Eq.(16)) for three intrinsic band-splitting scenarios under magnetic field modulation (Fig.~\ref{fig_lumi} (a)-(c)): (a) splitting in all bands, (b) splitting confined to conduction bands in both regions, and (c) splitting confined to valence bands in both regions, with fixed 0.1 eV  and Section III.A parameters. Fig.~\ref{fig_lumi} (d) reveals near-half polarization ($\vert P_I \vert \approx 50\%$) at $B =- 2B^*$ for asymmetric cases: scenario (b) yields dominant $\sigma^+$ emission, while (c) produces dominant $\sigma^-$ emission, whereas global splitting (a) shows cancellation. Crucially, high polarization persists at B=0 for (b) and (c) due to disparity between intrinsic conduction-band and intrinsic valence-band splitting respectively. This demonstrates the feasibility of achieving near $20\%$ circularly polarized emission without magnetic fields—a critical capability for spin-LEDs and quantum information interfaces. The circular polarization of this luminescence originates from the disparity between conduction-band and valence-band splitting in both P- and N-regions. When the splitting of conduction and valence bands are equal, the recombination rates for left- and right-circularly polarized light are identical. Conversely, when band splitting differs (such as in our calculated cases with only intrinsic conduction-band or intrinsic valence-band splitting), the recombination probabilities for opposite circular polarizations become asymmetric, thereby generating pronounced circularly polarized emission. As shown in Fig.~\ref{fig_current} (e) for the sensitivity of the circular polarization degree of luminescence to intrinsic conduction band splitting (case (b)) under magnetic field modulation, it can be seen that the intrinsic band splitting exceeds 0.1 eV, the magnetically-modulated circular polarization degree of luminescence stabilizes and ceases to be affected by the intrinsic splitting of the initial material. When the intrinsic splitting is small, the circular polarization degree of luminescence is still higher than the experimental results even without an external magnetic field. 
    \begin{figure*}[htbp!]
    	\centering
        \subfigure{\includegraphics[width=0.54\textwidth]{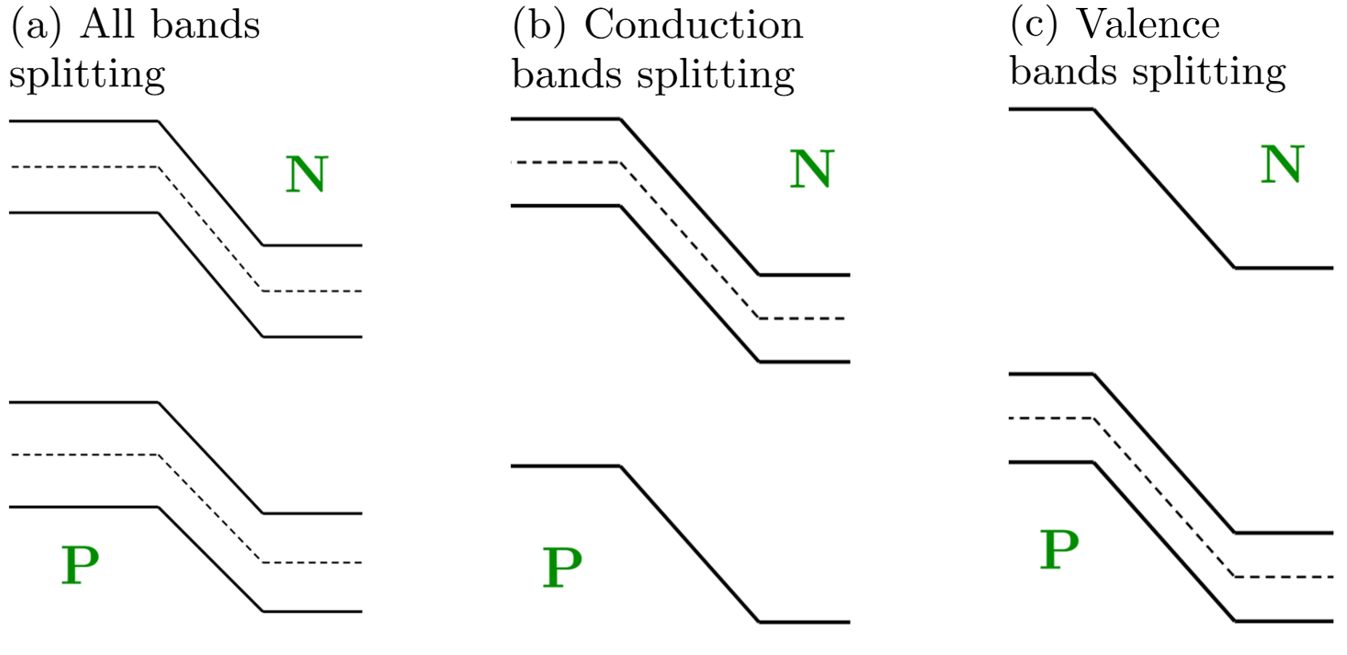}}\\
        \subfigure{\includegraphics[width=0.46\textwidth]{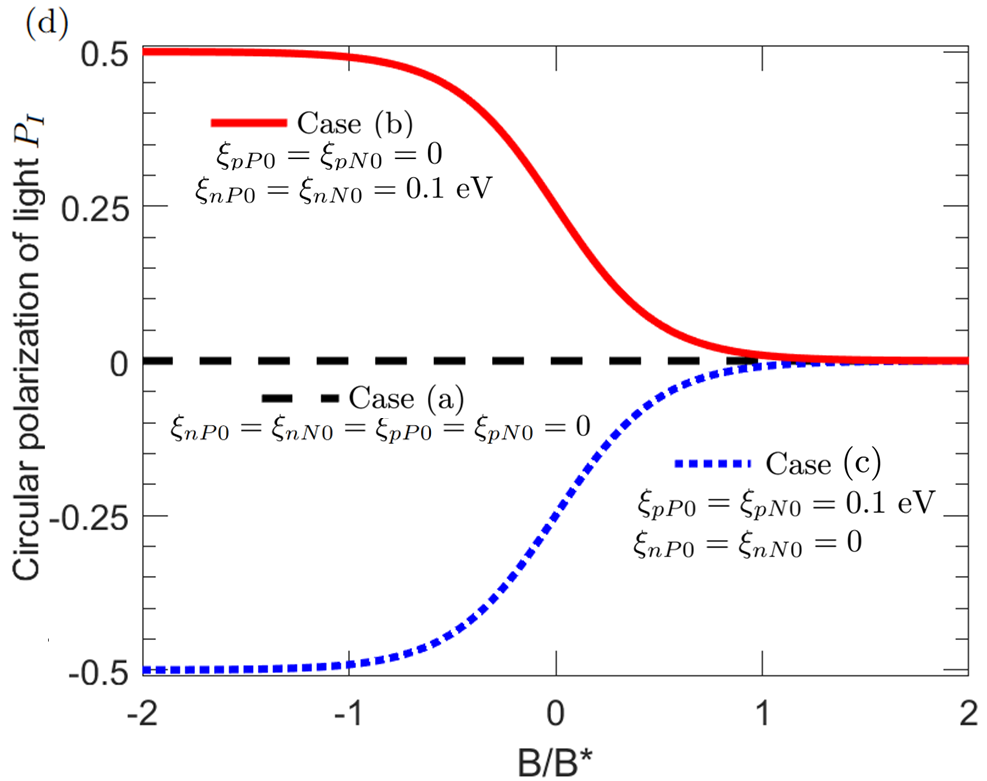}}
           \subfigure{\includegraphics[width=0.505\textwidth]{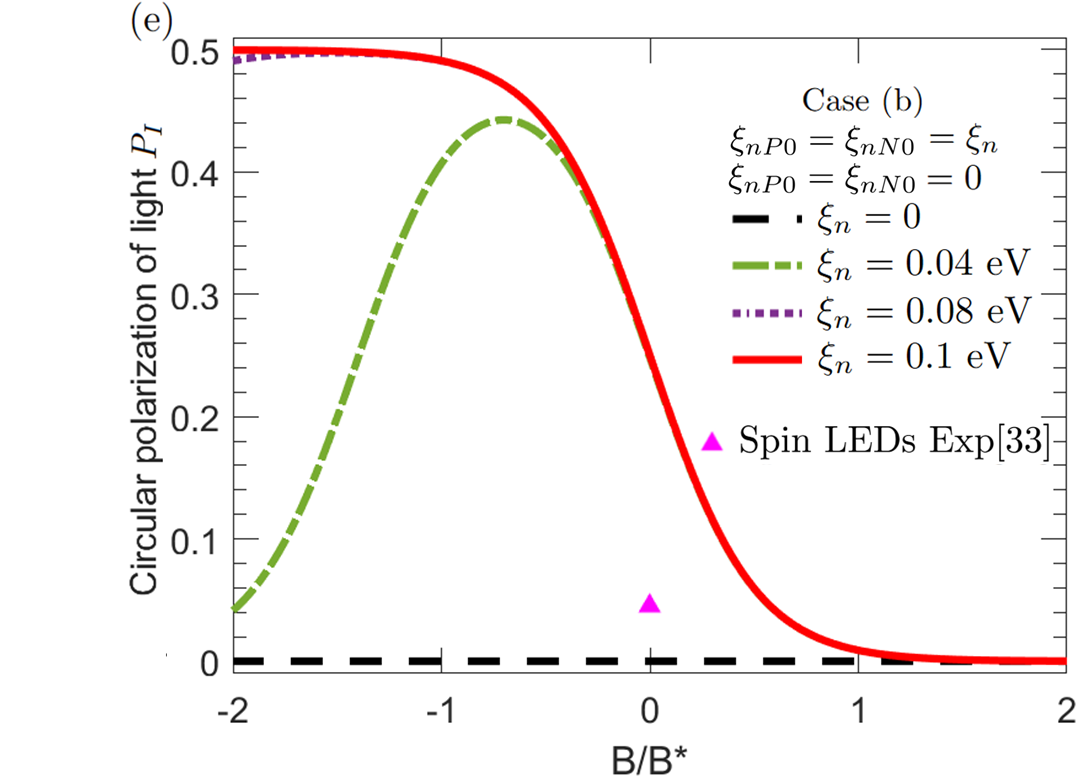}}
    
    	\caption{
        Magnetically tunable circular polarization in spin PN junctions via asymmetric band splitting. (a)-(c) Schematic band structures under zero external magnetic field for three scenarios: (a) Global splitting: Spin splitting in both conduction and valence bands of P and N regions. (b) Exclusive conduction-band splitting: Splitting confined to conduction bands in both regions. (c) Exclusive valence-band splitting: Splitting confined to valence bands in both regions. In all cases, intrinsic splitting $0.1\rm{eV}$ is fixed. (d) Circular polarization degree $P_I$ (Eq. (17)) as a function of magnetic field B (scaled to $B^*={2k_BT}/{g\mu_B}$) for each scenario. Case (a) (black dashed): Near-zero polarization due to cancellation between conduction- and valence-band splittings. Case (b) (red solid): Dominant $\sigma^+$ emission ($P_I\approx50\%$)  at $B=-2B^*$, driven by asymmetric recombination pathways for spin-conserving ($\uparrow\downarrow$) vs. spin-flip ($\uparrow\uparrow$) processes. Case (c) (blue dotted): Dominant $\sigma^-$ emission ($P_I\approx-50\%$) under the same conditions. (e) Sensitivity analysis: Circular polarization at $B=-2B^*$ for case (b) versus intrinsic conduction-band splitting $\xi_n$. Polarization saturates at $P_I>45\%$ for $\xi_n>0.1eV$ (purple marker: room-temperature spin-LED experimental reference). Physics insight: Angular-momentum-conserving recombination enables near-half polarization ($\vert P_I\vert\approx50\%$) when splitting is restricted to either conduction or valence bands, resolving spin-LED efficiency bottlenecks.
        }\label{fig_lumi}
    \end{figure*}

    \subsection{Spin Zener filter}
     \begin{figure*}[tphb]
    	\centering
         \subfigure{\includegraphics[width=0.30\textwidth]{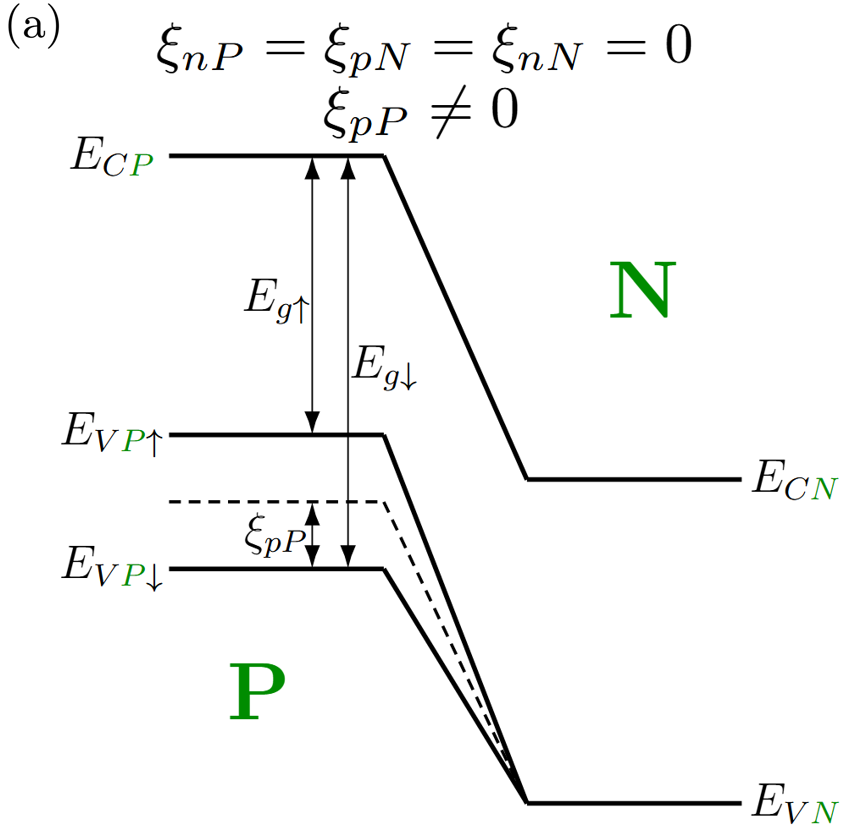}}\\
        \subfigure{\includegraphics[width=0.483\textwidth]{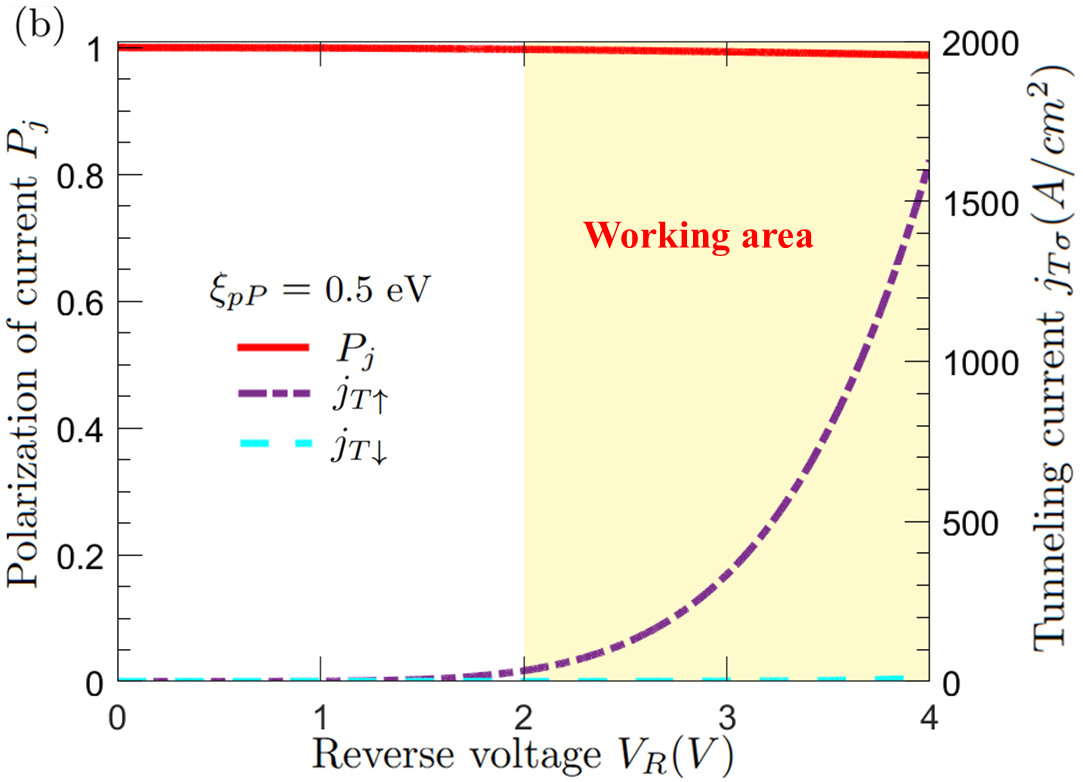}}
           \subfigure{\includegraphics[width=0.47\textwidth]{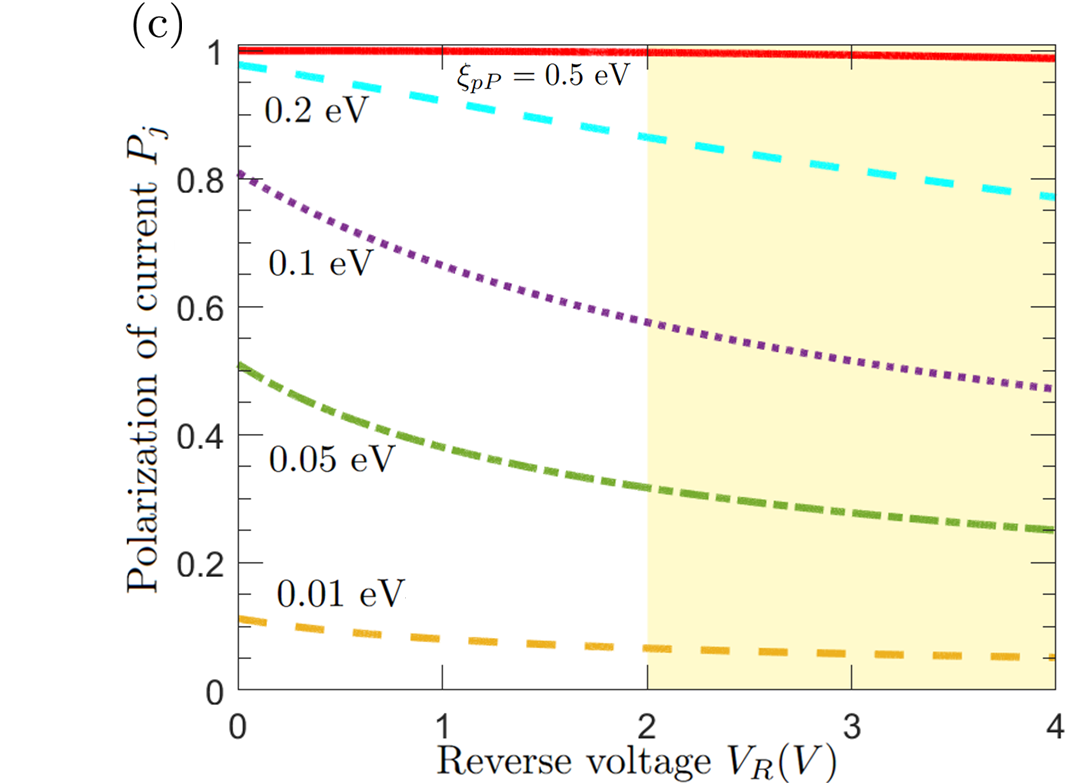}}
    	\caption{Spin Zener filter – Voltage-gated 100\% spin-polarized tunneling. (a) Band engineering principle: Schematic of reverse-biased spin PN junction under zero bias. Giant valence-band splitting in (Ga, Mn)As ($\xi_{pP}=0.5\rm{eV}$ from DFT ~\cite{16}) creates spin-dependent barrier heights ($ E_{g\uparrow}<E_{g\downarrow}$), enabling spin-resolved Zener tunneling. (b) Voltage-selective spin filter. Red curve: Current spin polarization $P_j$ (Eq. (28)) versus reverse bias $V_R$. Purple/blue curves: Spin-resolved tunneling currents $j_{T\uparrow}$ and $j_{T\downarrow}$ (Eqs. (26)-(27)) showing complete suppression of spin-down current. Critical insight: A voltage window $\triangle Vcrit\approx1V$ achieves $>99.9\%$ spin polarization – transforming conventional Zener breakdown into a spin filter. (c) Sensitivity analysis: Polarization $P_j$ at $V_R=\rm{2V-4V}$ versus valence-band splitting $\xi_{pP}$. Polarization exceeds 90\% for $\xi_{pP}>0.3\rm{eV}$ (yellow: operational regime where tunneling currents emerge). Parameters: $N_A=N_D=\rm{10^{19}cm^{-3}}$,$ V_{0\uparrow}=\rm{0.8V}$, $V_{0\downarrow}=\rm{1.8V}$, $\rm{\epsilon_{s}=13.1}$.
    	}\label{fig_Zenerfilter}
    \end{figure*}
	{We investigate $j_{T\sigma}$ reverse-bias spin-polarized transport in highly doped magnetic PN junctions, focusing on Zener breakdown, with the corresponding band diagram shown in  Fig.~\ref{fig_Zenerfilter} (a). In (Ga,Mn)As, large valence band spin splitting ($ \xi_{pP} \approx 0.5 \rm{eV}$, from DFT ~\cite{16}) creates spin-dependent built-in potentials: $V_{0\sigma}=V{_0}\pm\xi_{pP} (\sigma =\uparrow, \downarrow)$ with $V_{0\uparrow}=0.8V, V_{0\downarrow}=1.8V$, yielding spin-resolved electric fields under reverse bias $V_R$
    \begin{equation}
    E_\uparrow=\sqrt{\frac{2eN_AN_D(V_{0\uparrow}+V_R)}{\epsilon_s(N_A+N_D)}},
    \end{equation}
    \begin{equation}
    E_\downarrow=\sqrt{\frac{2eN_AN_D(V_{0\downarrow}+V_R)}{\epsilon_s(N_A+N_D)}}.
    \end{equation}
    For triangular barriers with spin-split band gaps $E_{g\sigma}$ (no conduction band splitting), the WKB tunneling probability
    \begin{equation}
    T_\uparrow=exp(-\frac{4\sqrt{2m_e^*}(E_{CP}-E_{VP\uparrow})^{3/2}}{3e\hbar E_\uparrow}),
    \end{equation}
    \begin{equation}
    T_\downarrow=exp(-\frac{4\sqrt{2m_e^*}(E_{CP}-E_{VP\downarrow})^{3/2}}{3e\hbar E_\downarrow}),
    \end{equation}
    and Kane-model currents~\cite{KANE1960181}
    \begin{equation}
    j_{T\uparrow}=\frac{\sqrt{2m_e^*}e^3E_{\uparrow}V_R }{8 \pi^2\hbar \sqrt{E_{g\uparrow}}}exp(-\frac{4\sqrt{2m_e^*}E_{g\uparrow}^{3/2}}{3e\hbar E_\uparrow}),
    \end{equation}
    \begin{equation}
    j_{T\downarrow}=\frac{\sqrt{2m_e^*}e^3E_{\downarrow}V_R }{8 \pi^2\hbar \sqrt{E_{g\downarrow}}}exp(-\frac{4\sqrt{2m_e^*}E_{g\downarrow}^{3/2}}{3e\hbar E_\downarrow}),
    \end{equation}
    give current polarization:
    \begin{equation}
    P_j=\frac{j_{T\uparrow}-j_{T\downarrow}}{j_{T\uparrow}+j_{T\downarrow}}.
    \end{equation}
    For the classic P type diluted magnetic semiconductor(DMS) (Ga, Mn)As, The doping of Mn impurities increases the hole concentration in GaAs to valence mismatch of $\rm{Mn}^{2+}$,which results in the low hole mobility about $\rm{10cm^2V^{-1}}$ at 300K for the hole concentration $n_h=10^{19}-10^{20}\rm{cm^{-3}}$ in (Ga,Mn)As~\cite{12,13,14}. Using parameters:  $N_A=N_D=10^{19}\rm{cm^{-3}}$, $\epsilon_s=13.1$, Fig.~\ref{fig_Zenerfilter}(b) shows near-unity spin polarization ($P_j \approx 100\%$) at critical voltage differences $\triangle V_{crit} =V_{crit\uparrow}-V_{crit\downarrow}  \approx 1 \rm{V}$. This demonstrates a voltage-controlled spin filter with perfect selectivity—providing new solution to the challenge of efficient spin injection without ferromagnetic contacts or complex heterostructures. The spin Zener filter effect arises because valence band splitting creates distinct breakdown thresholds, enabling complete suppression of one spin channel at tailored biases. As shown in Fig.~\ref{fig_Zenerfilter} (c) for the sensitivity of the spin Zener filter effect to the valence band splitting parameter in the P region, it can be seen that this effect is highly sensitive to valence band splitting in the P region. This stems from the fact that different valence band splittings correspond to different barrier differences for electrons of different spins, thereby affecting their tunneling probabilities (WKB model). It should be noted that significant tunneling current generally occurs only under reverse biases of 2-4 V. Consequently, the spin current polarization within this voltage range is meaningful. In our calculations, although higher polarization ratios occur at lower voltages, no significant tunneling current is present. The actual current polarization of tunnel current only becomes meaningful when exceeding breakdown voltage of one spin, as indicated by the
    working area in Fig.~\ref{fig_Zenerfilter}(b) and Fig.~\ref{fig_Zenerfilter}(c).

    \section{CONCLUSION}
    {
    This work establishes spin PN junctions—magnetic semiconductor homojunctions with spin-splitting-engineered band profiles—as a transformative platform for spintronic functionality by exploiting spin-dependent recombination probabilities ($r_{\sigma\sigma^\prime}=r_0\exp{\left([(E_{C\sigma}-E_{V{\sigma^\prime}})-(E_{C0}-E_{V0})]/k_BT\right)}$]). We resolve three foundational challenges: (i) achieving $100\times$ enhanced magnetoresistance sensitivity under forward bias via exponential modulation of recombination lifetimes by magnetic fields, overcoming sensitivity limits in conventional spintronics; (ii) enabling magnetically tunable near-half circular polarization ($|P_I| \approx 50\%$) in luminescence through angular-momentum-conserving recombination under asymmetric band splitting a critical advance for spin-LEDs; and (iii) demonstrating voltage-gated 100\% spin-polarized injection via the spin Zener filter, where giant valence-band splitting in (Ga,Mn)As creates spin-resolved Zener tunneling. These results unify transport amplification, polarization control, and spin injection within a single-junction architecture, extending Shockley’s framework to spin-resolved systems and establishing spin PN junctions as a universal design paradigm for CMOS-compatible quantum device engineering. 
    
   }

    \section {ACKNOWLEDGEMENTS}
    { This work is supported by National Key R\&D Pro
gram of China (Grant No. 2022YFA1405100), Chinese
 Academy of Sciences Project for Young Scientists in Ba
sic Research (Grant No. YSBR-030), and Basic Research
 Program of the Chinese Academy of Sciences Based on
 Major Scientific Infrastructures (Grant No. JZHKYPT
2021-08). GS was supported in part by the Innovation
 Program for Quantum Science and Technology under
 Grant No. 2024ZD0300500, NSFC No. 12447101 and the
 Strategic Priority Research Program of Chinese Academy
 of Sciences (Grant No. XDB1270000).
    }
    

\begin{thebibliography}{58}%
\makeatletter
\providecommand \@ifxundefined [1]{%
 \@ifx{#1\undefined}
}%
\providecommand \@ifnum [1]{%
 \ifnum #1\expandafter \@firstoftwo
 \else \expandafter \@secondoftwo
 \fi
}%
\providecommand \@ifx [1]{%
 \ifx #1\expandafter \@firstoftwo
 \else \expandafter \@secondoftwo
 \fi
}%
\providecommand \natexlab [1]{#1}%
\providecommand \enquote  [1]{``#1''}%
\providecommand \bibnamefont  [1]{#1}%
\providecommand \bibfnamefont [1]{#1}%
\providecommand \citenamefont [1]{#1}%
\providecommand \href@noop [0]{\@secondoftwo}%
\providecommand \href [0]{\begingroup \@sanitize@url \@href}%
\providecommand \@href[1]{\@@startlink{#1}\@@href}%
\providecommand \@@href[1]{\endgroup#1\@@endlink}%
\providecommand \@sanitize@url [0]{\catcode `\\12\catcode `\$12\catcode `\&12\catcode `\#12\catcode `\^12\catcode `\_12\catcode `\%12\relax}%
\providecommand \@@startlink[1]{}%
\providecommand \@@endlink[0]{}%
\providecommand \url  [0]{\begingroup\@sanitize@url \@url }%
\providecommand \@url [1]{\endgroup\@href {#1}{\urlprefix }}%
\providecommand \urlprefix  [0]{URL }%
\providecommand \Eprint [0]{\href }%
\providecommand \doibase [0]{https://doi.org/}%
\providecommand \selectlanguage [0]{\@gobble}%
\providecommand \bibinfo  [0]{\@secondoftwo}%
\providecommand \bibfield  [0]{\@secondoftwo}%
\providecommand \translation [1]{[#1]}%
\providecommand \BibitemOpen [0]{}%
\providecommand \bibitemStop [0]{}%
\providecommand \bibitemNoStop [0]{.\EOS\space}%
\providecommand \EOS [0]{\spacefactor3000\relax}%
\providecommand \BibitemShut  [1]{\csname bibitem#1\endcsname}%
\let\auto@bib@innerbib\@empty
\bibitem [{\citenamefont {\ifmmode \check{Z}\else \v{Z}\fi{}uti\ifmmode~\acute{c}\else \'{c}\fi{}}\ \emph {et~al.}(2004)\citenamefont {\ifmmode \check{Z}\else \v{Z}\fi{}uti\ifmmode~\acute{c}\else \'{c}\fi{}}, \citenamefont {Fabian},\ and\ \citenamefont {Das~Sarma}}]{RevModPhys.76.323}%
  \BibitemOpen
  \bibfield  {author} {\bibinfo {author} {\bibfnamefont {I.}~\bibnamefont {\ifmmode \check{Z}\else \v{Z}\fi{}uti\ifmmode~\acute{c}\else \'{c}\fi{}}}, \bibinfo {author} {\bibfnamefont {J.}~\bibnamefont {Fabian}},\ and\ \bibinfo {author} {\bibfnamefont {S.}~\bibnamefont {Das~Sarma}},\ }\bibfield  {title} {\bibinfo {title} {Spintronics: Fundamentals and applications},\ }\href {https://doi.org/10.1103/RevModPhys.76.323} {\bibfield  {journal} {\bibinfo  {journal} {Rev. Mod. Phys.}\ }\textbf {\bibinfo {volume} {76}},\ \bibinfo {pages} {323} (\bibinfo {year} {2004})}\BibitemShut {NoStop}%
\bibitem [{\citenamefont {Ohno}\ \emph {et~al.}(2000)\citenamefont {Ohno}, \citenamefont {Chiba}, \citenamefont {Matsukura}, \citenamefont {Omiya}, \citenamefont {Abe}, \citenamefont {Dietl}, \citenamefont {Ohno},\ and\ \citenamefont {Ohtani}}]{article}%
  \BibitemOpen
  \bibfield  {author} {\bibinfo {author} {\bibfnamefont {H.}~\bibnamefont {Ohno}}, \bibinfo {author} {\bibfnamefont {D.}~\bibnamefont {Chiba}}, \bibinfo {author} {\bibfnamefont {F.}~\bibnamefont {Matsukura}}, \bibinfo {author} {\bibfnamefont {T.}~\bibnamefont {Omiya}}, \bibinfo {author} {\bibfnamefont {E.}~\bibnamefont {Abe}}, \bibinfo {author} {\bibfnamefont {T.}~\bibnamefont {Dietl}}, \bibinfo {author} {\bibfnamefont {Y.}~\bibnamefont {Ohno}},\ and\ \bibinfo {author} {\bibfnamefont {K.}~\bibnamefont {Ohtani}},\ }\bibfield  {title} {\bibinfo {title} {Electric-field control of ferromagnetism},\ }\href {https://doi.org/10.1038/35050040} {\bibfield  {journal} {\bibinfo  {journal} {Nature}\ }\textbf {\bibinfo {volume} {408}},\ \bibinfo {pages} {944} (\bibinfo {year} {2000})}\BibitemShut {NoStop}%
\bibitem [{\citenamefont {Dietl}\ \emph {et~al.}(2000)\citenamefont {Dietl}, \citenamefont {Ohno}, \citenamefont {Matsukura}, \citenamefont {Cibert},\ and\ \citenamefont {Ferrand}}]{science}%
  \BibitemOpen
  \bibfield  {author} {\bibinfo {author} {\bibfnamefont {T.}~\bibnamefont {Dietl}}, \bibinfo {author} {\bibfnamefont {H.}~\bibnamefont {Ohno}}, \bibinfo {author} {\bibfnamefont {F.}~\bibnamefont {Matsukura}}, \bibinfo {author} {\bibfnamefont {J.}~\bibnamefont {Cibert}},\ and\ \bibinfo {author} {\bibfnamefont {D.}~\bibnamefont {Ferrand}},\ }\bibfield  {title} {\bibinfo {title} {Zener model description of ferromagnetism in zinc-blende magnetic semiconductors},\ }\href {https://doi.org/10.1126/science.287.5455.1019} {\bibfield  {journal} {\bibinfo  {journal} {Science}\ }\textbf {\bibinfo {volume} {287}},\ \bibinfo {pages} {1019} (\bibinfo {year} {2000})}\BibitemShut {NoStop}%
\bibitem [{\citenamefont {Awschalom}\ and\ \citenamefont {Flatté}(2007)}]{2}%
  \BibitemOpen
  \bibfield  {author} {\bibinfo {author} {\bibfnamefont {D.}~\bibnamefont {Awschalom}}\ and\ \bibinfo {author} {\bibfnamefont {M.}~\bibnamefont {Flatté}},\ }\bibfield  {title} {\bibinfo {title} {Challenges for semiconductor spintronics},\ }\href {https://doi.org/10.1038/nphys551} {\bibfield  {journal} {\bibinfo  {journal} {Nat. Phys.}\ }\textbf {\bibinfo {volume} {3}},\ \bibinfo {pages} {153} (\bibinfo {year} {2007})}\BibitemShut {NoStop}%
\bibitem [{\citenamefont {Matsukura}\ \emph {et~al.}(1998)\citenamefont {Matsukura}, \citenamefont {Ohno}, \citenamefont {Shen},\ and\ \citenamefont {Sugawara}}]{PhysRevB.57.R2037}%
  \BibitemOpen
  \bibfield  {author} {\bibinfo {author} {\bibfnamefont {F.}~\bibnamefont {Matsukura}}, \bibinfo {author} {\bibfnamefont {H.}~\bibnamefont {Ohno}}, \bibinfo {author} {\bibfnamefont {A.}~\bibnamefont {Shen}},\ and\ \bibinfo {author} {\bibfnamefont {Y.}~\bibnamefont {Sugawara}},\ }\bibfield  {title} {\bibinfo {title} {Transport properties and origin of ferromagnetism in \text{(Ga,Mn)As}},\ }\href {https://doi.org/10.1103/PhysRevB.57.R2037} {\bibfield  {journal} {\bibinfo  {journal} {Phys. Rev. B}\ }\textbf {\bibinfo {volume} {57}},\ \bibinfo {pages} {R2037} (\bibinfo {year} {1998})}\BibitemShut {NoStop}%
\bibitem [{\citenamefont {Mei}\ \emph {et~al.}(2023)\citenamefont {Mei}, \citenamefont {Tan}, \citenamefont {Cui}, \citenamefont {Wang}, \citenamefont {Yuan}, \citenamefont {Li}, \citenamefont {Lou}, \citenamefont {Hou}, \citenamefont {Zhao}, \citenamefont {Liu}, \citenamefont {Ji}, \citenamefont {Zhang}, \citenamefont {Feng},\ and\ \citenamefont {Cao}}]{MEI2023101251}%
  \BibitemOpen
  \bibfield  {author} {\bibinfo {author} {\bibfnamefont {G.}~\bibnamefont {Mei}}, \bibinfo {author} {\bibfnamefont {W.}~\bibnamefont {Tan}}, \bibinfo {author} {\bibfnamefont {X.}~\bibnamefont {Cui}}, \bibinfo {author} {\bibfnamefont {C.}~\bibnamefont {Wang}}, \bibinfo {author} {\bibfnamefont {Q.}~\bibnamefont {Yuan}}, \bibinfo {author} {\bibfnamefont {Y.}~\bibnamefont {Li}}, \bibinfo {author} {\bibfnamefont {C.}~\bibnamefont {Lou}}, \bibinfo {author} {\bibfnamefont {X.}~\bibnamefont {Hou}}, \bibinfo {author} {\bibfnamefont {M.}~\bibnamefont {Zhao}}, \bibinfo {author} {\bibfnamefont {Y.}~\bibnamefont {Liu}}, \bibinfo {author} {\bibfnamefont {W.}~\bibnamefont {Ji}}, \bibinfo {author} {\bibfnamefont {X.}~\bibnamefont {Zhang}}, \bibinfo {author} {\bibfnamefont {M.}~\bibnamefont {Feng}},\ and\ \bibinfo {author} {\bibfnamefont {L.}~\bibnamefont {Cao}},\ }\bibfield  {title} {\bibinfo {title} {Room-temperature ferromagnetism in \text{Fe}-doped \text{SnSe} bulk single crystalline semiconductor},\ }\href
  {https://doi.org/https://doi.org/10.1016/j.mtphys.2023.101251} {\bibfield  {journal} {\bibinfo  {journal} {Mater. Today Phys.}\ }\textbf {\bibinfo {volume} {38}},\ \bibinfo {pages} {101251} (\bibinfo {year} {2023})}\BibitemShut {NoStop}%
\bibitem [{\citenamefont {Chiba}\ \emph {et~al.}(2006)\citenamefont {Chiba}, \citenamefont {Matsukura},\ and\ \citenamefont {Ohno}}]{10.1063/1.2362971}%
  \BibitemOpen
  \bibfield  {author} {\bibinfo {author} {\bibfnamefont {D.}~\bibnamefont {Chiba}}, \bibinfo {author} {\bibfnamefont {F.}~\bibnamefont {Matsukura}},\ and\ \bibinfo {author} {\bibfnamefont {H.}~\bibnamefont {Ohno}},\ }\bibfield  {title} {\bibinfo {title} {Electric-field control of ferromagnetism in \text{(Ga,Mn)As}},\ }\href {https://doi.org/10.1063/1.2362971} {\bibfield  {journal} {\bibinfo  {journal} {Appl. Phys. Lett.}\ }\textbf {\bibinfo {volume} {89}},\ \bibinfo {pages} {162505} (\bibinfo {year} {2006})}\BibitemShut {NoStop}%
\bibitem [{\citenamefont {Jungwirth}\ \emph {et~al.}(2005)\citenamefont {Jungwirth}, \citenamefont {Wang}, \citenamefont {Ma\ifmmode~\check{s}\else \v{s}\fi{}ek}, \citenamefont {Edmonds}, \citenamefont {K\"onig}, \citenamefont {Sinova}, \citenamefont {Polini}, \citenamefont {Goncharuk}, \citenamefont {MacDonald}, \citenamefont {Sawicki}, \citenamefont {Rushforth}, \citenamefont {Campion}, \citenamefont {Zhao}, \citenamefont {Foxon},\ and\ \citenamefont {Gallagher}}]{PhysRevB.72.165204}%
  \BibitemOpen
  \bibfield  {author} {\bibinfo {author} {\bibfnamefont {T.}~\bibnamefont {Jungwirth}}, \bibinfo {author} {\bibfnamefont {K.~Y.}\ \bibnamefont {Wang}}, \bibinfo {author} {\bibfnamefont {J.}~\bibnamefont {Ma\ifmmode~\check{s}\else \v{s}\fi{}ek}}, \bibinfo {author} {\bibfnamefont {K.~W.}\ \bibnamefont {Edmonds}}, \bibinfo {author} {\bibfnamefont {J.}~\bibnamefont {K\"onig}}, \bibinfo {author} {\bibfnamefont {J.}~\bibnamefont {Sinova}}, \bibinfo {author} {\bibfnamefont {M.}~\bibnamefont {Polini}}, \bibinfo {author} {\bibfnamefont {N.~A.}\ \bibnamefont {Goncharuk}}, \bibinfo {author} {\bibfnamefont {A.~H.}\ \bibnamefont {MacDonald}}, \bibinfo {author} {\bibfnamefont {M.}~\bibnamefont {Sawicki}}, \bibinfo {author} {\bibfnamefont {A.~W.}\ \bibnamefont {Rushforth}}, \bibinfo {author} {\bibfnamefont {R.~P.}\ \bibnamefont {Campion}}, \bibinfo {author} {\bibfnamefont {L.~X.}\ \bibnamefont {Zhao}}, \bibinfo {author} {\bibfnamefont {C.~T.}\ \bibnamefont {Foxon}},\ and\ \bibinfo {author} {\bibfnamefont {B.~L.}\
  \bibnamefont {Gallagher}},\ }\bibfield  {title} {\bibinfo {title} {Prospects for high temperature ferromagnetism in \text{(Ga,Mn)As} semiconductors},\ }\href {https://doi.org/10.1103/PhysRevB.72.165204} {\bibfield  {journal} {\bibinfo  {journal} {Phys. Rev. B}\ }\textbf {\bibinfo {volume} {72}},\ \bibinfo {pages} {165204} (\bibinfo {year} {2005})}\BibitemShut {NoStop}%
\bibitem [{\citenamefont {Kudrin}\ \emph {et~al.}(2017)\citenamefont {Kudrin}, \citenamefont {Danilov}, \citenamefont {Lesnikov}, \citenamefont {Dorokhin}, \citenamefont {Vikhrova}, \citenamefont {Pavlov}, \citenamefont {Usov}, \citenamefont {Antonov}, \citenamefont {Kriukov}, \citenamefont {Alaferdov},\ and\ \citenamefont {Sobolev}}]{mag1}%
  \BibitemOpen
  \bibfield  {author} {\bibinfo {author} {\bibfnamefont {A.~V.}\ \bibnamefont {Kudrin}}, \bibinfo {author} {\bibfnamefont {Y.~A.}\ \bibnamefont {Danilov}}, \bibinfo {author} {\bibfnamefont {V.~P.}\ \bibnamefont {Lesnikov}}, \bibinfo {author} {\bibfnamefont {M.~V.}\ \bibnamefont {Dorokhin}}, \bibinfo {author} {\bibfnamefont {O.~V.}\ \bibnamefont {Vikhrova}}, \bibinfo {author} {\bibfnamefont {D.~A.}\ \bibnamefont {Pavlov}}, \bibinfo {author} {\bibfnamefont {Y.~V.}\ \bibnamefont {Usov}}, \bibinfo {author} {\bibfnamefont {I.~N.}\ \bibnamefont {Antonov}}, \bibinfo {author} {\bibfnamefont {R.~N.}\ \bibnamefont {Kriukov}}, \bibinfo {author} {\bibfnamefont {A.~V.}\ \bibnamefont {Alaferdov}},\ and\ \bibinfo {author} {\bibfnamefont {N.~A.}\ \bibnamefont {Sobolev}},\ }\bibfield  {title} {\bibinfo {title} {High-temperature intrinsic ferromagnetism in the \text{(In,Fe)Sb} semiconductor},\ }\href {https://doi.org/10.1063/1.5010191} {\bibfield  {journal} {\bibinfo  {journal} {J. Appl. Phys.}\ }\textbf {\bibinfo {volume}
  {122}},\ \bibinfo {pages} {183901} (\bibinfo {year} {2017})}\BibitemShut {NoStop}%
\bibitem [{\citenamefont {Ziebel}\ \emph {et~al.}(2024)\citenamefont {Ziebel}, \citenamefont {Feuer}, \citenamefont {Cox}, \citenamefont {Zhu}, \citenamefont {Dean},\ and\ \citenamefont {Roy}}]{mag2}%
  \BibitemOpen
  \bibfield  {author} {\bibinfo {author} {\bibfnamefont {M.~E.}\ \bibnamefont {Ziebel}}, \bibinfo {author} {\bibfnamefont {M.~L.}\ \bibnamefont {Feuer}}, \bibinfo {author} {\bibfnamefont {J.}~\bibnamefont {Cox}}, \bibinfo {author} {\bibfnamefont {X.}~\bibnamefont {Zhu}}, \bibinfo {author} {\bibfnamefont {C.~R.}\ \bibnamefont {Dean}},\ and\ \bibinfo {author} {\bibfnamefont {X.}~\bibnamefont {Roy}},\ }\bibfield  {title} {\bibinfo {title} {Crsbr: An air-stable, two-dimensional magnetic semiconductor},\ }\href {https://doi.org/10.1021/acs.nanolett.4c00624} {\bibfield  {journal} {\bibinfo  {journal} {Nano Lett.}\ }\textbf {\bibinfo {volume} {24}},\ \bibinfo {pages} {4319} (\bibinfo {year} {2024})}\BibitemShut {NoStop}%
\bibitem [{\citenamefont {Kobayashi}\ \emph {et~al.}(2021)\citenamefont {Kobayashi}, \citenamefont {Anh}, \citenamefont {Min\'ar}, \citenamefont {Khan}, \citenamefont {Borek}, \citenamefont {Hai}, \citenamefont {Harada}, \citenamefont {Schmitt}, \citenamefont {Oshima}, \citenamefont {Fujimori}, \citenamefont {Tanaka},\ and\ \citenamefont {Strocov}}]{mag3}%
  \BibitemOpen
  \bibfield  {author} {\bibinfo {author} {\bibfnamefont {M.}~\bibnamefont {Kobayashi}}, \bibinfo {author} {\bibfnamefont {L.~D.}\ \bibnamefont {Anh}}, \bibinfo {author} {\bibfnamefont {J.}~\bibnamefont {Min\'ar}}, \bibinfo {author} {\bibfnamefont {W.}~\bibnamefont {Khan}}, \bibinfo {author} {\bibfnamefont {S.}~\bibnamefont {Borek}}, \bibinfo {author} {\bibfnamefont {P.~N.}\ \bibnamefont {Hai}}, \bibinfo {author} {\bibfnamefont {Y.}~\bibnamefont {Harada}}, \bibinfo {author} {\bibfnamefont {T.}~\bibnamefont {Schmitt}}, \bibinfo {author} {\bibfnamefont {M.}~\bibnamefont {Oshima}}, \bibinfo {author} {\bibfnamefont {A.}~\bibnamefont {Fujimori}}, \bibinfo {author} {\bibfnamefont {M.}~\bibnamefont {Tanaka}},\ and\ \bibinfo {author} {\bibfnamefont {V.~N.}\ \bibnamefont {Strocov}},\ }\bibfield  {title} {\bibinfo {title} {Minority-spin impurity band in \text{$\mathrm{n}$-type (In,Fe)As}: A materials perspective for ferromagnetic semiconductors},\ }\href {https://doi.org/10.1103/PhysRevB.103.115111} {\bibfield
  {journal} {\bibinfo  {journal} {Phys. Rev. B}\ }\textbf {\bibinfo {volume} {103}},\ \bibinfo {pages} {115111} (\bibinfo {year} {2021})}\BibitemShut {NoStop}%
\bibitem [{\citenamefont {Vladimirova}\ \emph {et~al.}(2021)\citenamefont {Vladimirova}, \citenamefont {Scalbert}, \citenamefont {Kuznetsova},\ and\ \citenamefont {Kavokin}}]{mag4}%
  \BibitemOpen
  \bibfield  {author} {\bibinfo {author} {\bibfnamefont {M.}~\bibnamefont {Vladimirova}}, \bibinfo {author} {\bibfnamefont {D.}~\bibnamefont {Scalbert}}, \bibinfo {author} {\bibfnamefont {M.~S.}\ \bibnamefont {Kuznetsova}},\ and\ \bibinfo {author} {\bibfnamefont {K.~V.}\ \bibnamefont {Kavokin}},\ }\bibfield  {title} {\bibinfo {title} {Electron-induced nuclear magnetic ordering in $n$-type semiconductors},\ }\href {https://doi.org/10.1103/PhysRevB.103.205207} {\bibfield  {journal} {\bibinfo  {journal} {Phys. Rev. B}\ }\textbf {\bibinfo {volume} {103}},\ \bibinfo {pages} {205207} (\bibinfo {year} {2021})}\BibitemShut {NoStop}%
\bibitem [{\citenamefont {Fu}\ \emph {et~al.}(2021)\citenamefont {Fu}, \citenamefont {Gu}, \citenamefont {Zhi}, \citenamefont {Zhang}, \citenamefont {Zhang}, \citenamefont {Dong}, \citenamefont {Zhao}, \citenamefont {Xie},\ and\ \citenamefont {Ning}}]{mag5}%
  \BibitemOpen
  \bibfield  {author} {\bibinfo {author} {\bibfnamefont {L.}~\bibnamefont {Fu}}, \bibinfo {author} {\bibfnamefont {Y.}~\bibnamefont {Gu}}, \bibinfo {author} {\bibfnamefont {G.}~\bibnamefont {Zhi}}, \bibinfo {author} {\bibfnamefont {H.}~\bibnamefont {Zhang}}, \bibinfo {author} {\bibfnamefont {R.}~\bibnamefont {Zhang}}, \bibinfo {author} {\bibfnamefont {J.}~\bibnamefont {Dong}}, \bibinfo {author} {\bibfnamefont {X.}~\bibnamefont {Zhao}}, \bibinfo {author} {\bibfnamefont {L.}~\bibnamefont {Xie}},\ and\ \bibinfo {author} {\bibfnamefont {F.}~\bibnamefont {Ning}},\ }\bibfield  {title} {\bibinfo {title} {Drastic improvement of \text{Curie} temperature by chemical pressure in n-type diluted magnetic semiconductor \text{$\rm{Ba(Zn,Co)_2As_2}$}},\ }\href {https://api.semanticscholar.org/CorpusID:233185257} {\bibfield  {journal} {\bibinfo  {journal} {Sci. Rep.}\ }\textbf {\bibinfo {volume} {11 1}},\ \bibinfo {pages} {7652} (\bibinfo {year} {2021})}\BibitemShut {NoStop}%
\bibitem [{\citenamefont {You}\ \emph {et~al.}(2023)\citenamefont {You}, \citenamefont {Dong}, \citenamefont {Gu},\ and\ \citenamefont {Su}}]{cpl_40_6_067502}%
  \BibitemOpen
  \bibfield  {author} {\bibinfo {author} {\bibfnamefont {J.-Y.}\ \bibnamefont {You}}, \bibinfo {author} {\bibfnamefont {X.-J.}\ \bibnamefont {Dong}}, \bibinfo {author} {\bibfnamefont {B.}~\bibnamefont {Gu}},\ and\ \bibinfo {author} {\bibfnamefont {G.}~\bibnamefont {Su}},\ }\bibfield  {title} {\bibinfo {title} {Possible room-temperature ferromagnetic semiconductors},\ }\href {https://doi.org/10.1088/0256-307X/40/6/067502} {\bibfield  {journal} {\bibinfo  {journal} {Chin. Phys. Lett.}\ }\textbf {\bibinfo {volume} {40}},\ \bibinfo {pages} {067502} (\bibinfo {year} {2023})}\BibitemShut {NoStop}%
\bibitem [{\citenamefont {Datta}\ and\ \citenamefont {Das}(1990)}]{10.1063/1.102730}%
  \BibitemOpen
  \bibfield  {author} {\bibinfo {author} {\bibfnamefont {S.}~\bibnamefont {Datta}}\ and\ \bibinfo {author} {\bibfnamefont {B.}~\bibnamefont {Das}},\ }\bibfield  {title} {\bibinfo {title} {Electronic analog of the electro‐optic modulator},\ }\href {https://doi.org/10.1063/1.102730} {\bibfield  {journal} {\bibinfo  {journal} {Appl. Phys. Lett.}\ }\textbf {\bibinfo {volume} {56}},\ \bibinfo {pages} {665} (\bibinfo {year} {1990})}\BibitemShut {NoStop}%
\bibitem [{\citenamefont {Dyakonov}\ and\ \citenamefont {Perel}(1972)}]{3}%
  \BibitemOpen
  \bibfield  {author} {\bibinfo {author} {\bibfnamefont {M.}~\bibnamefont {Dyakonov}}\ and\ \bibinfo {author} {\bibfnamefont {V.}~\bibnamefont {Perel}},\ }\bibfield  {title} {\bibinfo {title} {Spin relaxation of conduction electrons in noncentrosymetric semiconductors},\ }\href@noop {} {\bibfield  {journal} {\bibinfo  {journal} {Solid State Phys.}\ }\textbf {\bibinfo {volume} {13}},\ \bibinfo {pages} {3023} (\bibinfo {year} {1972})}\BibitemShut {NoStop}%
\bibitem [{\citenamefont {Ohno}\ \emph {et~al.}(1996)\citenamefont {Ohno}, \citenamefont {Shen}, \citenamefont {Matsukura}, \citenamefont {Oiwa}, \citenamefont {Endo}, \citenamefont {Katsumoto},\ and\ \citenamefont {Iye}}]{4}%
  \BibitemOpen
  \bibfield  {author} {\bibinfo {author} {\bibfnamefont {H.}~\bibnamefont {Ohno}}, \bibinfo {author} {\bibfnamefont {A.}~\bibnamefont {Shen}}, \bibinfo {author} {\bibfnamefont {F.}~\bibnamefont {Matsukura}}, \bibinfo {author} {\bibfnamefont {A.}~\bibnamefont {Oiwa}}, \bibinfo {author} {\bibfnamefont {A.}~\bibnamefont {Endo}}, \bibinfo {author} {\bibfnamefont {S.}~\bibnamefont {Katsumoto}},\ and\ \bibinfo {author} {\bibfnamefont {Y.}~\bibnamefont {Iye}},\ }\bibfield  {title} {\bibinfo {title} {\text{(Ga,Mn)As}: A new diluted magnetic semiconductor based on \text{GaAs}},\ }\href {https://doi.org/10.1063/1.118061} {\bibfield  {journal} {\bibinfo  {journal} {Appl. Phys. Lett.}\ }\textbf {\bibinfo {volume} {69}},\ \bibinfo {pages} {363} (\bibinfo {year} {1996})}\BibitemShut {NoStop}%
\bibitem [{\citenamefont {Wolf}\ and\ \citenamefont {S.}(2001)}]{Wolf2001Spintronics}%
  \BibitemOpen
  \bibfield  {author} {\bibinfo {author} {\bibnamefont {Wolf}}\ and\ \bibinfo {author} {\bibfnamefont {A.}~\bibnamefont {S.}},\ }\bibfield  {title} {\bibinfo {title} {Spintronics: A spin-based electronics vision for the future},\ }\href@noop {} {\bibfield  {journal} {\bibinfo  {journal} {Science}\ }\textbf {\bibinfo {volume} {294}},\ \bibinfo {pages} {1488} (\bibinfo {year} {2001})}\BibitemShut {NoStop}%
\bibitem [{\citenamefont {Chappert}\ \emph {et~al.}(2007)\citenamefont {Chappert}, \citenamefont {Albert},\ and\ \citenamefont {Nguyen-Van-Dau}}]{2014The}%
  \BibitemOpen
  \bibfield  {author} {\bibinfo {author} {\bibfnamefont {C.}~\bibnamefont {Chappert}}, \bibinfo {author} {\bibfnamefont {F.}~\bibnamefont {Albert}},\ and\ \bibinfo {author} {\bibfnamefont {F.}~\bibnamefont {Nguyen-Van-Dau}},\ }\bibfield  {title} {\bibinfo {title} {The emergence of spin electronics in data storage},\ }\href {https://doi.org/10.1038/nmat2024} {\bibfield  {journal} {\bibinfo  {journal} {Nat. Mater.}\ }\textbf {\bibinfo {volume} {6}},\ \bibinfo {pages} {813} (\bibinfo {year} {2007})}\BibitemShut {NoStop}%
\bibitem [{\citenamefont {Garello}\ \emph {et~al.}(2019)\citenamefont {Garello}, \citenamefont {Yasin},\ and\ \citenamefont {Kar}}]{8739466}%
  \BibitemOpen
  \bibfield  {author} {\bibinfo {author} {\bibfnamefont {K.}~\bibnamefont {Garello}}, \bibinfo {author} {\bibfnamefont {F.}~\bibnamefont {Yasin}},\ and\ \bibinfo {author} {\bibfnamefont {G.~S.}\ \bibnamefont {Kar}},\ }\bibfield  {title} {\bibinfo {title} {Spin-orbit torque \text{MRAM} for ultrafast embedded memories: from fundamentals to large scale technology integration},\ }in\ \href {https://doi.org/10.1109/IMW.2019.8739466} {\emph {\bibinfo {booktitle} {Proc. IEEE Int. Mem. Workshop (IMW)}}}\ (\bibinfo {year} {2019})\ pp.\ \bibinfo {pages} {1--4}\BibitemShut {NoStop}%
\bibitem [{\citenamefont {Hirohata}\ \emph {et~al.}(2015)\citenamefont {Hirohata}, \citenamefont {Sukegawa}, \citenamefont {Yanagihara}, \citenamefont {Žutić}, \citenamefont {Seki}, \citenamefont {Mizukami},\ and\ \citenamefont {Swaminathan}}]{spinDevice}%
  \BibitemOpen
  \bibfield  {author} {\bibinfo {author} {\bibfnamefont {A.}~\bibnamefont {Hirohata}}, \bibinfo {author} {\bibfnamefont {H.}~\bibnamefont {Sukegawa}}, \bibinfo {author} {\bibfnamefont {H.}~\bibnamefont {Yanagihara}}, \bibinfo {author} {\bibfnamefont {I.}~\bibnamefont {Žutić}}, \bibinfo {author} {\bibfnamefont {T.}~\bibnamefont {Seki}}, \bibinfo {author} {\bibfnamefont {S.}~\bibnamefont {Mizukami}},\ and\ \bibinfo {author} {\bibfnamefont {R.}~\bibnamefont {Swaminathan}},\ }\bibfield  {title} {\bibinfo {title} {Roadmap for emerging materials for spintronic device applications},\ }\href {https://doi.org/10.1109/TMAG.2015.2457393} {\bibfield  {journal} {\bibinfo  {journal} {IEEE Trans. Magn.}\ }\textbf {\bibinfo {volume} {51}},\ \bibinfo {pages} {1} (\bibinfo {year} {2015})}\BibitemShut {NoStop}%
\bibitem [{\citenamefont {Tao}\ \emph {et~al.}(2020)\citenamefont {Tao}, \citenamefont {Naeemi},\ and\ \citenamefont {Tsymbal}}]{Spingate}%
  \BibitemOpen
  \bibfield  {author} {\bibinfo {author} {\bibfnamefont {L.~L.}\ \bibnamefont {Tao}}, \bibinfo {author} {\bibfnamefont {A.}~\bibnamefont {Naeemi}},\ and\ \bibinfo {author} {\bibfnamefont {E.~Y.}\ \bibnamefont {Tsymbal}},\ }\bibfield  {title} {\bibinfo {title} {Valley-spin logic gates},\ }\href {https://doi.org/10.1103/PhysRevApplied.13.054043} {\bibfield  {journal} {\bibinfo  {journal} {Phys. Rev. Appl.}\ }\textbf {\bibinfo {volume} {13}},\ \bibinfo {pages} {054043} (\bibinfo {year} {2020})}\BibitemShut {NoStop}%
\bibitem [{\citenamefont {Kajale}\ \emph {et~al.}(2024)\citenamefont {Kajale}, \citenamefont {Nguyễn}, \citenamefont {Chao}, \citenamefont {Bono}, \citenamefont {Boonkird}, \citenamefont {Li},\ and\ \citenamefont {Sarkar}}]{5}%
  \BibitemOpen
  \bibfield  {author} {\bibinfo {author} {\bibfnamefont {S.}~\bibnamefont {Kajale}}, \bibinfo {author} {\bibfnamefont {T.}~\bibnamefont {Nguyen}}, \bibinfo {author} {\bibfnamefont {C.}~\bibnamefont {Chao}}, \bibinfo {author} {\bibfnamefont {D.}~\bibnamefont {Bono}}, \bibinfo {author} {\bibfnamefont {A.}~\bibnamefont {Boonkird}}, \bibinfo {author} {\bibfnamefont {M.}~\bibnamefont {Li}},\ and\ \bibinfo {author} {\bibfnamefont {D.}~\bibnamefont {Sarkar}},\ }\bibfield  {title} {\bibinfo {title} {Current-induced switching of a \text{van der Waals} ferromagnet at room temperature},\ }\href {https://doi.org/10.1038/s41467-024-45586-4} {\bibfield  {journal} {\bibinfo  {journal} {Nat. Commun.}\ }\textbf {\bibinfo {volume} {15}},\ \bibinfo {pages} {1485} (\bibinfo {year} {2024})}\BibitemShut {NoStop}%
\bibitem [{\citenamefont {Guillet}\ \emph {et~al.}(2020)\citenamefont {Guillet}, \citenamefont {Zucchetti}, \citenamefont {Marchionni}, \citenamefont {Hallal}, \citenamefont {Biagioni}, \citenamefont {Vergnaud}, \citenamefont {Marty}, \citenamefont {Okuno}, \citenamefont {Masseboeuf}, \citenamefont {Finazzi}, \citenamefont {Ciccacci}, \citenamefont {Chshiev}, \citenamefont {Bottegoni},\ and\ \citenamefont {Jamet}}]{6}%
  \BibitemOpen
  \bibfield  {author} {\bibinfo {author} {\bibfnamefont {T.}~\bibnamefont {Guillet}}, \bibinfo {author} {\bibfnamefont {C.}~\bibnamefont {Zucchetti}}, \bibinfo {author} {\bibfnamefont {A.}~\bibnamefont {Marchionni}}, \bibinfo {author} {\bibfnamefont {A.}~\bibnamefont {Hallal}}, \bibinfo {author} {\bibfnamefont {P.}~\bibnamefont {Biagioni}}, \bibinfo {author} {\bibfnamefont {C.}~\bibnamefont {Vergnaud}}, \bibinfo {author} {\bibfnamefont {A.}~\bibnamefont {Marty}}, \bibinfo {author} {\bibfnamefont {H.}~\bibnamefont {Okuno}}, \bibinfo {author} {\bibfnamefont {A.}~\bibnamefont {Masseboeuf}}, \bibinfo {author} {\bibfnamefont {M.}~\bibnamefont {Finazzi}}, \bibinfo {author} {\bibfnamefont {F.}~\bibnamefont {Ciccacci}}, \bibinfo {author} {\bibfnamefont {M.}~\bibnamefont {Chshiev}}, \bibinfo {author} {\bibfnamefont {F.}~\bibnamefont {Bottegoni}},\ and\ \bibinfo {author} {\bibfnamefont {M.}~\bibnamefont {Jamet}},\ }\bibfield  {title} {\bibinfo {title} {Spin orbitronics at a topological insulator-semiconductor
  interface},\ }\href {https://doi.org/10.1103/PhysRevB.101.184406} {\bibfield  {journal} {\bibinfo  {journal} {Phys. Rev. B}\ }\textbf {\bibinfo {volume} {101}},\ \bibinfo {pages} {184406} (\bibinfo {year} {2020})}\BibitemShut {NoStop}%
\bibitem [{\citenamefont {Ilan}\ \emph {et~al.}(2015)\citenamefont {Ilan}, \citenamefont {de~Juan},\ and\ \citenamefont {Moore}}]{topoPN0}%
  \BibitemOpen
  \bibfield  {author} {\bibinfo {author} {\bibfnamefont {R.}~\bibnamefont {Ilan}}, \bibinfo {author} {\bibfnamefont {F.}~\bibnamefont {de~Juan}},\ and\ \bibinfo {author} {\bibfnamefont {J.~E.}\ \bibnamefont {Moore}},\ }\bibfield  {title} {\bibinfo {title} {Spin-based mach-zehnder interferometry in topological insulator p-n junctions},\ }\href {https://doi.org/10.1103/PhysRevLett.115.096802} {\bibfield  {journal} {\bibinfo  {journal} {Phys. Rev. Lett.}\ }\textbf {\bibinfo {volume} {115}},\ \bibinfo {pages} {096802} (\bibinfo {year} {2015})}\BibitemShut {NoStop}%
\bibitem [{\citenamefont {Wang}\ \emph {et~al.}(2012)\citenamefont {Wang}, \citenamefont {Chen}, \citenamefont {Zhu},\ and\ \citenamefont {Zhang}}]{topoPN}%
  \BibitemOpen
  \bibfield  {author} {\bibinfo {author} {\bibfnamefont {J.}~\bibnamefont {Wang}}, \bibinfo {author} {\bibfnamefont {X.}~\bibnamefont {Chen}}, \bibinfo {author} {\bibfnamefont {B.-F.}\ \bibnamefont {Zhu}},\ and\ \bibinfo {author} {\bibfnamefont {S.-C.}\ \bibnamefont {Zhang}},\ }\bibfield  {title} {\bibinfo {title} {Topological p-n junction},\ }\href {https://doi.org/10.1103/PhysRevB.85.235131} {\bibfield  {journal} {\bibinfo  {journal} {Phys. Rev. B}\ }\textbf {\bibinfo {volume} {85}},\ \bibinfo {pages} {235131} (\bibinfo {year} {2012})}\BibitemShut {NoStop}%
\bibitem [{\citenamefont {{Otani}}\ \emph {et~al.}(2017)\citenamefont {{Otani}}, \citenamefont {{Shiraishi}}, \citenamefont {{Oiwa}}, \citenamefont {{Saitoh}},\ and\ \citenamefont {{Murakami}}}]{topo1}%
  \BibitemOpen
  \bibfield  {author} {\bibinfo {author} {\bibfnamefont {Y.}~\bibnamefont {{Otani}}}, \bibinfo {author} {\bibfnamefont {M.}~\bibnamefont {{Shiraishi}}}, \bibinfo {author} {\bibfnamefont {A.}~\bibnamefont {{Oiwa}}}, \bibinfo {author} {\bibfnamefont {E.}~\bibnamefont {{Saitoh}}},\ and\ \bibinfo {author} {\bibfnamefont {S.}~\bibnamefont {{Murakami}}},\ }\bibfield  {title} {\bibinfo {title} {{Spin conversion on the nanoscale}},\ }\href {https://doi.org/10.1038/nphys4192} {\bibfield  {journal} {\bibinfo  {journal} {Nat. Phys.}\ }\textbf {\bibinfo {volume} {13}},\ \bibinfo {pages} {829} (\bibinfo {year} {2017})}\BibitemShut {NoStop}%
\bibitem [{\citenamefont {Soumyanarayanan}\ \emph {et~al.}(2016)\citenamefont {Soumyanarayanan}, \citenamefont {Reyren}, \citenamefont {Albert},\ and\ \citenamefont {Panagopoulos}}]{topo2}%
  \BibitemOpen
  \bibfield  {author} {\bibinfo {author} {\bibfnamefont {A.}~\bibnamefont {Soumyanarayanan}}, \bibinfo {author} {\bibfnamefont {N.}~\bibnamefont {Reyren}}, \bibinfo {author} {\bibfnamefont {F.}~\bibnamefont {Albert}},\ and\ \bibinfo {author} {\bibfnamefont {C.}~\bibnamefont {Panagopoulos}},\ }\bibfield  {title} {\bibinfo {title} {Emergent phenomena induced by spin-orbit coupling at surfaces and interfaces},\ }\href {https://doi.org/10.1038/nature19820} {\bibfield  {journal} {\bibinfo  {journal} {Nature}\ }\textbf {\bibinfo {volume} {539}},\ \bibinfo {pages} {509} (\bibinfo {year} {2016})}\BibitemShut {NoStop}%
\bibitem [{\citenamefont {Chen}\ \emph {et~al.}(2006)\citenamefont {Chen}, \citenamefont {Moser}, \citenamefont {Kotissek}, \citenamefont {Sadowski}, \citenamefont {Zenger}, \citenamefont {Weiss},\ and\ \citenamefont {Wegscheider}}]{electricalspin}%
  \BibitemOpen
  \bibfield  {author} {\bibinfo {author} {\bibfnamefont {P.}~\bibnamefont {Chen}}, \bibinfo {author} {\bibfnamefont {J.}~\bibnamefont {Moser}}, \bibinfo {author} {\bibfnamefont {P.}~\bibnamefont {Kotissek}}, \bibinfo {author} {\bibfnamefont {J.}~\bibnamefont {Sadowski}}, \bibinfo {author} {\bibfnamefont {M.}~\bibnamefont {Zenger}}, \bibinfo {author} {\bibfnamefont {D.}~\bibnamefont {Weiss}},\ and\ \bibinfo {author} {\bibfnamefont {W.}~\bibnamefont {Wegscheider}},\ }\bibfield  {title} {\bibinfo {title} {All electrical measurement of spin injection in a magnetic p-n junction diode},\ }\href {https://doi.org/10.1103/PhysRevB.74.241302} {\bibfield  {journal} {\bibinfo  {journal} {Phys. Rev. B}\ }\textbf {\bibinfo {volume} {74}},\ \bibinfo {pages} {241302} (\bibinfo {year} {2006})}\BibitemShut {NoStop}%
\bibitem [{\citenamefont {Komatsu}\ \emph {et~al.}(2014)\citenamefont {Komatsu}, \citenamefont {Kasai}, \citenamefont {Li}, \citenamefont {Nakaharai}, \citenamefont {Mitoma}, \citenamefont {Yamamoto},\ and\ \citenamefont {Tsukagoshi}}]{SpinExp1}%
  \BibitemOpen
  \bibfield  {author} {\bibinfo {author} {\bibfnamefont {K.}~\bibnamefont {Komatsu}}, \bibinfo {author} {\bibfnamefont {S.}~\bibnamefont {Kasai}}, \bibinfo {author} {\bibfnamefont {S.}~\bibnamefont {Li}}, \bibinfo {author} {\bibfnamefont {S.}~\bibnamefont {Nakaharai}}, \bibinfo {author} {\bibfnamefont {N.}~\bibnamefont {Mitoma}}, \bibinfo {author} {\bibfnamefont {M.}~\bibnamefont {Yamamoto}},\ and\ \bibinfo {author} {\bibfnamefont {K.}~\bibnamefont {Tsukagoshi}},\ }\bibfield  {title} {\bibinfo {title} {Spin injection and detection in a graphene lateral spin valve using an yttrium-oxide tunneling barrier},\ }\href {https://doi.org/10.7567/APEX.7.085101} {\bibfield  {journal} {\bibinfo  {journal} {Appl. Phys. Express}\ }\textbf {\bibinfo {volume} {7}},\ \bibinfo {pages} {085101} (\bibinfo {year} {2014})}\BibitemShut {NoStop}%
\bibitem [{\citenamefont {Liang}\ \emph {et~al.}(2017)\citenamefont {Liang}, \citenamefont {Yang}, \citenamefont {Renucci}, \citenamefont {Tao}, \citenamefont {Laczkowski}, \citenamefont {Mc-Murtry}, \citenamefont {Wang}, \citenamefont {Marie}, \citenamefont {George}, \citenamefont {Petit-Watelot} \emph {et~al.}}]{SpinExp2}%
  \BibitemOpen
  \bibfield  {author} {\bibinfo {author} {\bibfnamefont {S.}~\bibnamefont {Liang}}, \bibinfo {author} {\bibfnamefont {H.}~\bibnamefont {Yang}}, \bibinfo {author} {\bibfnamefont {P.}~\bibnamefont {Renucci}}, \bibinfo {author} {\bibfnamefont {B.}~\bibnamefont {Tao}}, \bibinfo {author} {\bibfnamefont {P.}~\bibnamefont {Laczkowski}}, \bibinfo {author} {\bibfnamefont {S.}~\bibnamefont {Mc-Murtry}}, \bibinfo {author} {\bibfnamefont {G.}~\bibnamefont {Wang}}, \bibinfo {author} {\bibfnamefont {X.}~\bibnamefont {Marie}}, \bibinfo {author} {\bibfnamefont {J.-M.}\ \bibnamefont {George}}, \bibinfo {author} {\bibfnamefont {S.}~\bibnamefont {Petit-Watelot}}, \emph {et~al.},\ }\bibfield  {title} {\bibinfo {title} {Electrical spin injection and detection in molybdenum disulfide multilayer channel},\ }\href {https://doi.org/10.1038/ncomms14947} {\bibfield  {journal} {\bibinfo  {journal} {Nat. Commun.}\ }\textbf {\bibinfo {volume} {8}},\ \bibinfo {pages} {14947} (\bibinfo {year} {2017})}\BibitemShut {NoStop}%
\bibitem [{\citenamefont {Holub}\ and\ \citenamefont {Bhattacharya}(2007)}]{Holub_2007}%
  \BibitemOpen
  \bibfield  {author} {\bibinfo {author} {\bibfnamefont {M.}~\bibnamefont {Holub}}\ and\ \bibinfo {author} {\bibfnamefont {P.}~\bibnamefont {Bhattacharya}},\ }\bibfield  {title} {\bibinfo {title} {Spin-polarized light-emitting diodes and lasers},\ }\href {https://doi.org/10.1088/0022-3727/40/11/R01} {\bibfield  {journal} {\bibinfo  {journal} {J. Phys. D: Appl. Phys.}\ }\textbf {\bibinfo {volume} {40}},\ \bibinfo {pages} {R179} (\bibinfo {year} {2007})}\BibitemShut {NoStop}%
\bibitem [{\citenamefont {Kim}\ \emph {et~al.}(2021)\citenamefont {Kim}, \citenamefont {Zhai}, \citenamefont {Lu}, \citenamefont {Pan}, \citenamefont {Xiao}, \citenamefont {Gaulding}, \citenamefont {Harvey}, \citenamefont {Berry}, \citenamefont {Vardeny}, \citenamefont {Luther},\ and\ \citenamefont {Beard}}]{LED1}%
  \BibitemOpen
  \bibfield  {author} {\bibinfo {author} {\bibfnamefont {Y.-H.}\ \bibnamefont {Kim}}, \bibinfo {author} {\bibfnamefont {Y.}~\bibnamefont {Zhai}}, \bibinfo {author} {\bibfnamefont {H.}~\bibnamefont {Lu}}, \bibinfo {author} {\bibfnamefont {X.}~\bibnamefont {Pan}}, \bibinfo {author} {\bibfnamefont {C.}~\bibnamefont {Xiao}}, \bibinfo {author} {\bibfnamefont {E.~A.}\ \bibnamefont {Gaulding}}, \bibinfo {author} {\bibfnamefont {S.~P.}\ \bibnamefont {Harvey}}, \bibinfo {author} {\bibfnamefont {J.~J.}\ \bibnamefont {Berry}}, \bibinfo {author} {\bibfnamefont {Z.~V.}\ \bibnamefont {Vardeny}}, \bibinfo {author} {\bibfnamefont {J.~M.}\ \bibnamefont {Luther}},\ and\ \bibinfo {author} {\bibfnamefont {M.~C.}\ \bibnamefont {Beard}},\ }\bibfield  {title} {\bibinfo {title} {Chiral-induced spin selectivity enables a room-temperature spin light-emitting diode},\ }\href {https://doi.org/10.1126/science.abf5291} {\bibfield  {journal} {\bibinfo  {journal} {Science}\ }\textbf {\bibinfo {volume} {371}},\ \bibinfo {pages} {1129}
  (\bibinfo {year} {2021})}\BibitemShut {NoStop}%
\bibitem [{\citenamefont {Li}\ \emph {et~al.}(2025{\natexlab{a}})\citenamefont {Li}, \citenamefont {Li}, \citenamefont {Yuan}, \citenamefont {Zhang}, \citenamefont {Tao}, \citenamefont {Zhan}, \citenamefont {Yu}, \citenamefont {Wang}, \citenamefont {liu}, \citenamefont {Wang},\ and\ \citenamefont {Qin}}]{LED2}%
  \BibitemOpen
  \bibfield  {author} {\bibinfo {author} {\bibfnamefont {B.}~\bibnamefont {Li}}, \bibinfo {author} {\bibfnamefont {Y.}~\bibnamefont {Li}}, \bibinfo {author} {\bibfnamefont {W.}~\bibnamefont {Yuan}}, \bibinfo {author} {\bibfnamefont {X.}~\bibnamefont {Zhang}}, \bibinfo {author} {\bibfnamefont {S.}~\bibnamefont {Tao}}, \bibinfo {author} {\bibfnamefont {H.}~\bibnamefont {Zhan}}, \bibinfo {author} {\bibfnamefont {Z.-g.}\ \bibnamefont {Yu}}, \bibinfo {author} {\bibfnamefont {K.}~\bibnamefont {Wang}}, \bibinfo {author} {\bibfnamefont {J.}~\bibnamefont {liu}}, \bibinfo {author} {\bibfnamefont {L.}~\bibnamefont {Wang}},\ and\ \bibinfo {author} {\bibfnamefont {C.}~\bibnamefont {Qin}},\ }\bibfield  {title} {\bibinfo {title} {Chiral quasi-\text{2D} perovskites based single junction spin-light-emitting diodes},\ }\href {https://doi.org/https://doi.org/10.1002/adfm.202415433} {\bibfield  {journal} {\bibinfo  {journal} {Adv. Funct. Mater.}\ }\textbf {\bibinfo {volume} {35}},\ \bibinfo {pages} {2415433} (\bibinfo {year}
  {2025}{\natexlab{a}})}\BibitemShut {NoStop}%
\bibitem [{\citenamefont {Yao}\ \emph {et~al.}(2025)\citenamefont {Yao}, \citenamefont {Huang}, \citenamefont {Sun}, \citenamefont {Wang}, \citenamefont {Xue}, \citenamefont {Huang}, \citenamefont {Dong}, \citenamefont {Chen},\ and\ \citenamefont {Lu}}]{LED3}%
  \BibitemOpen
  \bibfield  {author} {\bibinfo {author} {\bibfnamefont {J.}~\bibnamefont {Yao}}, \bibinfo {author} {\bibfnamefont {Y.}~\bibnamefont {Huang}}, \bibinfo {author} {\bibfnamefont {H.}~\bibnamefont {Sun}}, \bibinfo {author} {\bibfnamefont {Z.}~\bibnamefont {Wang}}, \bibinfo {author} {\bibfnamefont {J.}~\bibnamefont {Xue}}, \bibinfo {author} {\bibfnamefont {Z.}~\bibnamefont {Huang}}, \bibinfo {author} {\bibfnamefont {S.-C.}\ \bibnamefont {Dong}}, \bibinfo {author} {\bibfnamefont {X.}~\bibnamefont {Chen}},\ and\ \bibinfo {author} {\bibfnamefont {H.}~\bibnamefont {Lu}},\ }\bibfield  {title} {\bibinfo {title} {Efficient spin-light-emitting diodes with tunable red to near-infrared emission at room temperature},\ }\href {https://doi.org/https://doi.org/10.1002/adma.202413669} {\bibfield  {journal} {\bibinfo  {journal} {Adv. Mater.}\ }\textbf {\bibinfo {volume} {37}},\ \bibinfo {pages} {2413669} (\bibinfo {year} {2025})}\BibitemShut {NoStop}%
\bibitem [{\citenamefont {Han}\ \emph {et~al.}(2017)\citenamefont {Han}, \citenamefont {Richardella}, \citenamefont {Siddiqui}, \citenamefont {Finley}, \citenamefont {Samarth},\ and\ \citenamefont {Liu}}]{7}%
  \BibitemOpen
  \bibfield  {author} {\bibinfo {author} {\bibfnamefont {J.}~\bibnamefont {Han}}, \bibinfo {author} {\bibfnamefont {A.}~\bibnamefont {Richardella}}, \bibinfo {author} {\bibfnamefont {S.~A.}\ \bibnamefont {Siddiqui}}, \bibinfo {author} {\bibfnamefont {J.}~\bibnamefont {Finley}}, \bibinfo {author} {\bibfnamefont {N.}~\bibnamefont {Samarth}},\ and\ \bibinfo {author} {\bibfnamefont {L.}~\bibnamefont {Liu}},\ }\bibfield  {title} {\bibinfo {title} {Room-temperature spin-orbit torque switching induced by a topological insulator},\ }\href {https://doi.org/10.1103/PhysRevLett.119.077702} {\bibfield  {journal} {\bibinfo  {journal} {Phys. Rev. Lett.}\ }\textbf {\bibinfo {volume} {119}},\ \bibinfo {pages} {077702} (\bibinfo {year} {2017})}\BibitemShut {NoStop}%
\bibitem [{\citenamefont {\ifmmode \check{Z}\else \v{Z}\fi{}uti\ifmmode~\acute{c}\else \'{c}\fi{}}\ \emph {et~al.}(2001)\citenamefont {\ifmmode \check{Z}\else \v{Z}\fi{}uti\ifmmode~\acute{c}\else \'{c}\fi{}}, \citenamefont {Fabian},\ and\ \citenamefont {Das~Sarma}}]{PhysRevB.64.121201}%
  \BibitemOpen
  \bibfield  {author} {\bibinfo {author} {\bibfnamefont {I.}~\bibnamefont {\ifmmode \check{Z}\else \v{Z}\fi{}uti\ifmmode~\acute{c}\else \'{c}\fi{}}}, \bibinfo {author} {\bibfnamefont {J.}~\bibnamefont {Fabian}},\ and\ \bibinfo {author} {\bibfnamefont {S.}~\bibnamefont {Das~Sarma}},\ }\bibfield  {title} {\bibinfo {title} {Spin injection through the depletion layer: A theory of spin-polarized p-n junctions and solar cells},\ }\href {https://doi.org/10.1103/PhysRevB.64.121201} {\bibfield  {journal} {\bibinfo  {journal} {Phys. Rev. B}\ }\textbf {\bibinfo {volume} {64}},\ \bibinfo {pages} {121201} (\bibinfo {year} {2001})}\BibitemShut {NoStop}%
\bibitem [{\citenamefont {\ifmmode \check{Z}\else \v{Z}\fi{}uti\ifmmode~\acute{c}\else \'{c}\fi{}}\ \emph {et~al.}(2002)\citenamefont {\ifmmode \check{Z}\else \v{Z}\fi{}uti\ifmmode~\acute{c}\else \'{c}\fi{}}, \citenamefont {Fabian},\ and\ \citenamefont {Das~Sarma}}]{PhysRevLett.88.066603}%
  \BibitemOpen
  \bibfield  {author} {\bibinfo {author} {\bibfnamefont {I.}~\bibnamefont {\ifmmode \check{Z}\else \v{Z}\fi{}uti\ifmmode~\acute{c}\else \'{c}\fi{}}}, \bibinfo {author} {\bibfnamefont {J.}~\bibnamefont {Fabian}},\ and\ \bibinfo {author} {\bibfnamefont {S.}~\bibnamefont {Das~Sarma}},\ }\bibfield  {title} {\bibinfo {title} {Spin-polarized transport in inhomogeneous magnetic semiconductors: Theory of magnetic/nonmagnetic p-n junctions},\ }\href {https://doi.org/10.1103/PhysRevLett.88.066603} {\bibfield  {journal} {\bibinfo  {journal} {Phys. Rev. Lett.}\ }\textbf {\bibinfo {volume} {88}},\ \bibinfo {pages} {066603} (\bibinfo {year} {2002})}\BibitemShut {NoStop}%
\bibitem [{\citenamefont {Fabian}\ \emph {et~al.}(2002)\citenamefont {Fabian}, \citenamefont {\ifmmode \check{Z}\else \v{Z}\fi{}uti\ifmmode~\acute{c}\else \'{c}\fi{}},\ and\ \citenamefont {Das~Sarma}}]{PhysRevB.66.165301}%
  \BibitemOpen
  \bibfield  {author} {\bibinfo {author} {\bibfnamefont {J.}~\bibnamefont {Fabian}}, \bibinfo {author} {\bibfnamefont {I.}~\bibnamefont {\ifmmode \check{Z}\else \v{Z}\fi{}uti\ifmmode~\acute{c}\else \'{c}\fi{}}},\ and\ \bibinfo {author} {\bibfnamefont {S.}~\bibnamefont {Das~Sarma}},\ }\bibfield  {title} {\bibinfo {title} {Theory of spin-polarized bipolar transport in magnetic p-n junctions},\ }\href {https://doi.org/10.1103/PhysRevB.66.165301} {\bibfield  {journal} {\bibinfo  {journal} {Phys. Rev. B}\ }\textbf {\bibinfo {volume} {66}},\ \bibinfo {pages} {165301} (\bibinfo {year} {2002})}\BibitemShut {NoStop}%
\bibitem [{\citenamefont {\ifmmode \check{Z}\else \v{Z}\fi{}uti\ifmmode~\acute{c}\else \'{c}\fi{}}\ \emph {et~al.}(2006)\citenamefont {\ifmmode \check{Z}\else \v{Z}\fi{}uti\ifmmode~\acute{c}\else \'{c}\fi{}}, \citenamefont {Fabian},\ and\ \citenamefont {Erwin}}]{PhysRevLett.97.026602}%
  \BibitemOpen
  \bibfield  {author} {\bibinfo {author} {\bibfnamefont {I.}~\bibnamefont {\ifmmode \check{Z}\else \v{Z}\fi{}uti\ifmmode~\acute{c}\else \'{c}\fi{}}}, \bibinfo {author} {\bibfnamefont {J.}~\bibnamefont {Fabian}},\ and\ \bibinfo {author} {\bibfnamefont {S.~C.}\ \bibnamefont {Erwin}},\ }\bibfield  {title} {\bibinfo {title} {Spin injection and detection in silicon},\ }\href {https://doi.org/10.1103/PhysRevLett.97.026602} {\bibfield  {journal} {\bibinfo  {journal} {Phys. Rev. Lett.}\ }\textbf {\bibinfo {volume} {97}},\ \bibinfo {pages} {026602} (\bibinfo {year} {2006})}\BibitemShut {NoStop}%
\bibitem [{\citenamefont {Yamakage}\ \emph {et~al.}(2009)\citenamefont {Yamakage}, \citenamefont {Imura}, \citenamefont {Cayssol},\ and\ \citenamefont {Kuramoto}}]{grePN}%
  \BibitemOpen
  \bibfield  {author} {\bibinfo {author} {\bibfnamefont {A.}~\bibnamefont {Yamakage}}, \bibinfo {author} {\bibfnamefont {K.-I.}\ \bibnamefont {Imura}}, \bibinfo {author} {\bibfnamefont {J.}~\bibnamefont {Cayssol}},\ and\ \bibinfo {author} {\bibfnamefont {Y.}~\bibnamefont {Kuramoto}},\ }\bibfield  {title} {\bibinfo {title} {Spin-orbit effects in a graphene bipolar p-n junction},\ }\href {https://doi.org/10.1209/0295-5075/87/47005} {\bibfield  {journal} {\bibinfo  {journal} {Europhys. Lett.}\ }\textbf {\bibinfo {volume} {87}},\ \bibinfo {pages} {47005} (\bibinfo {year} {2009})}\BibitemShut {NoStop}%
\bibitem [{\citenamefont {Bercioux}\ and\ \citenamefont {De~Martino}(2019)}]{SOCgphenePN}%
  \BibitemOpen
  \bibfield  {author} {\bibinfo {author} {\bibfnamefont {D.}~\bibnamefont {Bercioux}}\ and\ \bibinfo {author} {\bibfnamefont {A.}~\bibnamefont {De~Martino}},\ }\bibfield  {title} {\bibinfo {title} {Spin-orbit interaction and snake states in a graphene p-n junction},\ }\href {https://doi.org/10.1103/PhysRevB.100.115407} {\bibfield  {journal} {\bibinfo  {journal} {Phys. Rev. B}\ }\textbf {\bibinfo {volume} {100}},\ \bibinfo {pages} {115407} (\bibinfo {year} {2019})}\BibitemShut {NoStop}%
\bibitem [{\citenamefont {Pacuski}\ \emph {et~al.}(2006)\citenamefont {Pacuski}, \citenamefont {Ferrand}, \citenamefont {Kossacki}, \citenamefont {Marcet}, \citenamefont {Cibert}, \citenamefont {Gaj},\ and\ \citenamefont {Golnik}}]{8}%
  \BibitemOpen
  \bibfield  {author} {\bibinfo {author} {\bibfnamefont {W.}~\bibnamefont {Pacuski}}, \bibinfo {author} {\bibfnamefont {D.}~\bibnamefont {Ferrand}}, \bibinfo {author} {\bibfnamefont {P.}~\bibnamefont {Kossacki}}, \bibinfo {author} {\bibfnamefont {S.}~\bibnamefont {Marcet}}, \bibinfo {author} {\bibfnamefont {J.}~\bibnamefont {Cibert}}, \bibinfo {author} {\bibfnamefont {J.}~\bibnamefont {Gaj}},\ and\ \bibinfo {author} {\bibfnamefont {A.}~\bibnamefont {Golnik}},\ }\bibfield  {title} {\bibinfo {title} {Excitonic giant \text{Zeeman} effect in wide gap diluted magnetic semiconductors based on \text{ZnO} and \text{GaN}},\ }\href {https://doi.org/10.12693/APhysPolA.110.303} {\bibfield  {journal} {\bibinfo  {journal} {Acta Phys. Pol. A}\ }\textbf {\bibinfo {volume} {110}} (\bibinfo {year} {2006})}\BibitemShut {NoStop}%
\bibitem [{\citenamefont {Dietl}\ and\ \citenamefont {Ohno}(2014)}]{9}%
  \BibitemOpen
  \bibfield  {author} {\bibinfo {author} {\bibfnamefont {T.}~\bibnamefont {Dietl}}\ and\ \bibinfo {author} {\bibfnamefont {H.}~\bibnamefont {Ohno}},\ }\bibfield  {title} {\bibinfo {title} {Dilute ferromagnetic semiconductors: Physics and spintronic structures},\ }\href {https://doi.org/10.1103/RevModPhys.86.187} {\bibfield  {journal} {\bibinfo  {journal} {Rev. Mod. Phys.}\ }\textbf {\bibinfo {volume} {86}},\ \bibinfo {pages} {187} (\bibinfo {year} {2014})}\BibitemShut {NoStop}%
\bibitem [{\citenamefont {Zener}\ and\ \citenamefont {Fowler}(1934)}]{10}%
  \BibitemOpen
  \bibfield  {author} {\bibinfo {author} {\bibfnamefont {C.}~\bibnamefont {Zener}}\ and\ \bibinfo {author} {\bibfnamefont {R.~H.}\ \bibnamefont {Fowler}},\ }\bibfield  {title} {\bibinfo {title} {A theory of the electrical breakdown of solid dielectrics},\ }\href {https://doi.org/10.1098/rspa.1934.0116} {\bibfield  {journal} {\bibinfo  {journal} {Proc. R. Soc. Lond. A}\ }\textbf {\bibinfo {volume} {145}},\ \bibinfo {pages} {523} (\bibinfo {year} {1934})}\BibitemShut {NoStop}%
\bibitem [{\citenamefont {Esaki}(1958)}]{11}%
  \BibitemOpen
  \bibfield  {author} {\bibinfo {author} {\bibfnamefont {L.}~\bibnamefont {Esaki}},\ }\bibfield  {title} {\bibinfo {title} {New phenomenon in narrow \text{Germanium} p-n junctions},\ }\href {https://doi.org/10.1103/PhysRev.109.603} {\bibfield  {journal} {\bibinfo  {journal} {Phys. Rev.}\ }\textbf {\bibinfo {volume} {109}},\ \bibinfo {pages} {603} (\bibinfo {year} {1958})}\BibitemShut {NoStop}%
\bibitem [{\citenamefont {Kohda}\ \emph {et~al.}(2001)\citenamefont {Kohda}, \citenamefont {Ohno}, \citenamefont {Takamura}, \citenamefont {Matsukura},\ and\ \citenamefont {Ohno}}]{SPINESAKI}%
  \BibitemOpen
  \bibfield  {author} {\bibinfo {author} {\bibfnamefont {M.}~\bibnamefont {Kohda}}, \bibinfo {author} {\bibfnamefont {Y.}~\bibnamefont {Ohno}}, \bibinfo {author} {\bibfnamefont {K.}~\bibnamefont {Takamura}}, \bibinfo {author} {\bibfnamefont {F.}~\bibnamefont {Matsukura}},\ and\ \bibinfo {author} {\bibfnamefont {H.}~\bibnamefont {Ohno}},\ }\bibfield  {title} {\bibinfo {title} {A spin \text{Esaki} diode},\ }\href {https://doi.org/10.1143/JJAP.40.L1274} {\bibfield  {journal} {\bibinfo  {journal} {Jpn. J. Appl. Phys.}\ }\textbf {\bibinfo {volume} {40}},\ \bibinfo {pages} {L1274} (\bibinfo {year} {2001})}\BibitemShut {NoStop}%
\bibitem [{Sup()}]{Supply}%
  \BibitemOpen
  \bibfield  {title} {\bibinfo {title} {See \text{Supplemental Material} at [url will be inserted by publisher] for additional details of equations},\ }\href@noop {} {\ }\BibitemShut {NoStop}%
\bibitem [{\citenamefont {Shockley}(1949)}]{15}%
  \BibitemOpen
  \bibfield  {author} {\bibinfo {author} {\bibfnamefont {W.}~\bibnamefont {Shockley}},\ }\bibfield  {title} {\bibinfo {title} {The theory of p-n junctions in semiconductors and p-n junction transistors},\ }\href {https://doi.org/10.1002/j.1538-7305.1949.tb03645.x} {\bibfield  {journal} {\bibinfo  {journal} {Bell Syst. Tech. J.}\ }\textbf {\bibinfo {volume} {28}},\ \bibinfo {pages} {435} (\bibinfo {year} {1949})}\BibitemShut {NoStop}%
\bibitem [{\citenamefont {Lei}\ \emph {et~al.}(2020)\citenamefont {Lei}, \citenamefont {Lehner}, \citenamefont {Rubi}, \citenamefont {Cheah}, \citenamefont {Karalic}, \citenamefont {Mittag}, \citenamefont {Alt}, \citenamefont {Scharnetzky}, \citenamefont {M\"arki}, \citenamefont {Zeitler}, \citenamefont {Wegscheider}, \citenamefont {Ihn},\ and\ \citenamefont {Ensslin}}]{gf1}%
  \BibitemOpen
  \bibfield  {author} {\bibinfo {author} {\bibfnamefont {Z.}~\bibnamefont {Lei}}, \bibinfo {author} {\bibfnamefont {C.~A.}\ \bibnamefont {Lehner}}, \bibinfo {author} {\bibfnamefont {K.}~\bibnamefont {Rubi}}, \bibinfo {author} {\bibfnamefont {E.}~\bibnamefont {Cheah}}, \bibinfo {author} {\bibfnamefont {M.}~\bibnamefont {Karalic}}, \bibinfo {author} {\bibfnamefont {C.}~\bibnamefont {Mittag}}, \bibinfo {author} {\bibfnamefont {L.}~\bibnamefont {Alt}}, \bibinfo {author} {\bibfnamefont {J.}~\bibnamefont {Scharnetzky}}, \bibinfo {author} {\bibfnamefont {P.}~\bibnamefont {M\"arki}}, \bibinfo {author} {\bibfnamefont {U.}~\bibnamefont {Zeitler}}, \bibinfo {author} {\bibfnamefont {W.}~\bibnamefont {Wegscheider}}, \bibinfo {author} {\bibfnamefont {T.}~\bibnamefont {Ihn}},\ and\ \bibinfo {author} {\bibfnamefont {K.}~\bibnamefont {Ensslin}},\ }\bibfield  {title} {\bibinfo {title} {Electronic $g$ factor and magnetotransport in \text{InSb} quantum wells},\ }\href {https://doi.org/10.1103/PhysRevResearch.2.033213} {\bibfield
   {journal} {\bibinfo  {journal} {Phys. Rev. Res.}\ }\textbf {\bibinfo {volume} {2}},\ \bibinfo {pages} {033213} (\bibinfo {year} {2020})}\BibitemShut {NoStop}%
\bibitem [{\citenamefont {Jiang}\ \emph {et~al.}(2022)\citenamefont {Jiang}, \citenamefont {Ermolaev}, \citenamefont {Kipshidze}, \citenamefont {Moon}, \citenamefont {Ozerov}, \citenamefont {Smirnov}, \citenamefont {Jiang},\ and\ \citenamefont {Suchalkin}}]{gf2}%
  \BibitemOpen
  \bibfield  {author} {\bibinfo {author} {\bibfnamefont {Y.}~\bibnamefont {Jiang}}, \bibinfo {author} {\bibfnamefont {M.}~\bibnamefont {Ermolaev}}, \bibinfo {author} {\bibfnamefont {G.}~\bibnamefont {Kipshidze}}, \bibinfo {author} {\bibfnamefont {S.}~\bibnamefont {Moon}}, \bibinfo {author} {\bibfnamefont {M.}~\bibnamefont {Ozerov}}, \bibinfo {author} {\bibfnamefont {D.}~\bibnamefont {Smirnov}}, \bibinfo {author} {\bibfnamefont {Z.}~\bibnamefont {Jiang}},\ and\ \bibinfo {author} {\bibfnamefont {S.}~\bibnamefont {Suchalkin}},\ }\bibfield  {title} {\bibinfo {title} {Giant g-factors and fully spin-polarized states in metamorphic short-period \text{InAsSb/InSb} superlattices},\ }\href {https://doi.org/10.1038/s41467-022-33560-x} {\bibfield  {journal} {\bibinfo  {journal} {Nat. Commun.}\ }\textbf {\bibinfo {volume} {13}} (\bibinfo {year} {2022})}\BibitemShut {NoStop}%
\bibitem [{\citenamefont {Tu}\ \emph {et~al.}(2019)\citenamefont {Tu}, \citenamefont {Hai}, \citenamefont {Anh},\ and\ \citenamefont {Tanaka}}]{Tu_2019}%
  \BibitemOpen
  \bibfield  {author} {\bibinfo {author} {\bibfnamefont {N.~T.}\ \bibnamefont {Tu}}, \bibinfo {author} {\bibfnamefont {P.~N.}\ \bibnamefont {Hai}}, \bibinfo {author} {\bibfnamefont {L.~D.}\ \bibnamefont {Anh}},\ and\ \bibinfo {author} {\bibfnamefont {M.}~\bibnamefont {Tanaka}},\ }\bibfield  {title} {\bibinfo {title} {Heavily \text{Fe}-doped ferromagnetic semiconductor \text{(In,Fe)Sb} with high \text{Curie} temperature and large magnetic anisotropy},\ }\href {https://doi.org/10.7567/1882-0786/ab3f4b} {\bibfield  {journal} {\bibinfo  {journal} {Appl. Phys. Express}\ }\textbf {\bibinfo {volume} {12}},\ \bibinfo {pages} {103004} (\bibinfo {year} {2019})}\BibitemShut {NoStop}%
\bibitem [{\citenamefont {You}\ \emph {et~al.}(2020)\citenamefont {You}, \citenamefont {Gu}, \citenamefont {Maekawa},\ and\ \citenamefont {Su}}]{PhysRevB.102.094432}%
  \BibitemOpen
  \bibfield  {author} {\bibinfo {author} {\bibfnamefont {J.-Y.}\ \bibnamefont {You}}, \bibinfo {author} {\bibfnamefont {B.}~\bibnamefont {Gu}}, \bibinfo {author} {\bibfnamefont {S.}~\bibnamefont {Maekawa}},\ and\ \bibinfo {author} {\bibfnamefont {G.}~\bibnamefont {Su}},\ }\bibfield  {title} {\bibinfo {title} {Microscopic mechanism of high-temperature ferromagnetism in \text{Fe, Mn, and Cr-doped InSb, InAs, and GaSb} magnetic semiconductors},\ }\href {https://doi.org/10.1103/PhysRevB.102.094432} {\bibfield  {journal} {\bibinfo  {journal} {Phys. Rev. B}\ }\textbf {\bibinfo {volume} {102}},\ \bibinfo {pages} {094432} (\bibinfo {year} {2020})}\BibitemShut {NoStop}%
\bibitem [{\citenamefont {Li}\ \emph {et~al.}(2025{\natexlab{b}})\citenamefont {Li}, \citenamefont {Li}, \citenamefont {You}, \citenamefont {Su},\ and\ \citenamefont {Gu}}]{16}%
  \BibitemOpen
  \bibfield  {author} {\bibinfo {author} {\bibfnamefont {X.}~\bibnamefont {Li}}, \bibinfo {author} {\bibfnamefont {J.-W.}\ \bibnamefont {Li}}, \bibinfo {author} {\bibfnamefont {J.-Y.}\ \bibnamefont {You}}, \bibinfo {author} {\bibfnamefont {G.}~\bibnamefont {Su}},\ and\ \bibinfo {author} {\bibfnamefont {B.}~\bibnamefont {Gu}},\ }\bibfield  {title} {\bibinfo {title} {High \text{Curie} temperature in diluted magnetic semiconductors \text{(B, Mn)$X$ ($X=\text{N}$, P, As, Sb)}},\ }\href {https://doi.org/10.1103/PhysRevB.111.184425} {\bibfield  {journal} {\bibinfo  {journal} {Phys. Rev. B}\ }\textbf {\bibinfo {volume} {111}},\ \bibinfo {pages} {184425} (\bibinfo {year} {2025}{\natexlab{b}})}\BibitemShut {NoStop}%
\bibitem [{\citenamefont {Kane}(1960)}]{KANE1960181}%
  \BibitemOpen
  \bibfield  {author} {\bibinfo {author} {\bibfnamefont {E.}~\bibnamefont {Kane}},\ }\bibfield  {title} {\bibinfo {title} {Zener tunneling in semiconductors},\ }\href {https://doi.org/https://doi.org/10.1016/0022-3697(60)90035-4} {\bibfield  {journal} {\bibinfo  {journal} {J. Phys. Chem. Solids}\ }\textbf {\bibinfo {volume} {12}},\ \bibinfo {pages} {181} (\bibinfo {year} {1960})}\BibitemShut {NoStop}%
\bibitem [{\citenamefont {Limmer}\ \emph {et~al.}(2002)\citenamefont {Limmer}, \citenamefont {Glunk}, \citenamefont {Mascheck}, \citenamefont {Koeder}, \citenamefont {Klarer}, \citenamefont {Schoch}, \citenamefont {Thonke}, \citenamefont {Sauer},\ and\ \citenamefont {Waag}}]{12}%
  \BibitemOpen
  \bibfield  {author} {\bibinfo {author} {\bibfnamefont {W.}~\bibnamefont {Limmer}}, \bibinfo {author} {\bibfnamefont {M.}~\bibnamefont {Glunk}}, \bibinfo {author} {\bibfnamefont {S.}~\bibnamefont {Mascheck}}, \bibinfo {author} {\bibfnamefont {A.}~\bibnamefont {Koeder}}, \bibinfo {author} {\bibfnamefont {D.}~\bibnamefont {Klarer}}, \bibinfo {author} {\bibfnamefont {W.}~\bibnamefont {Schoch}}, \bibinfo {author} {\bibfnamefont {K.}~\bibnamefont {Thonke}}, \bibinfo {author} {\bibfnamefont {R.}~\bibnamefont {Sauer}},\ and\ \bibinfo {author} {\bibfnamefont {A.}~\bibnamefont {Waag}},\ }\bibfield  {title} {\bibinfo {title} {Coupled plasmon--\text{LO}-phonon modes in \text{${\mathrm{Ga}}_{1\ensuremath{-}x}{\mathrm{Mn}}_{x}\mathrm{As}$}},\ }\href {https://doi.org/10.1103/PhysRevB.66.205209} {\bibfield  {journal} {\bibinfo  {journal} {Phys. Rev. B}\ }\textbf {\bibinfo {volume} {66}},\ \bibinfo {pages} {205209} (\bibinfo {year} {2002})}\BibitemShut {NoStop}%
\bibitem [{\citenamefont {Limmer}\ \emph {et~al.}(2005)\citenamefont {Limmer}, \citenamefont {Koeder}, \citenamefont {Frank}, \citenamefont {Avrutin}, \citenamefont {Schoch}, \citenamefont {Sauer}, \citenamefont {Zuern}, \citenamefont {Eisenmenger}, \citenamefont {Ziemann}, \citenamefont {Peiner},\ and\ \citenamefont {Waag}}]{13}%
  \BibitemOpen
  \bibfield  {author} {\bibinfo {author} {\bibfnamefont {W.}~\bibnamefont {Limmer}}, \bibinfo {author} {\bibfnamefont {A.}~\bibnamefont {Koeder}}, \bibinfo {author} {\bibfnamefont {S.}~\bibnamefont {Frank}}, \bibinfo {author} {\bibfnamefont {V.}~\bibnamefont {Avrutin}}, \bibinfo {author} {\bibfnamefont {W.}~\bibnamefont {Schoch}}, \bibinfo {author} {\bibfnamefont {R.}~\bibnamefont {Sauer}}, \bibinfo {author} {\bibfnamefont {K.}~\bibnamefont {Zuern}}, \bibinfo {author} {\bibfnamefont {J.}~\bibnamefont {Eisenmenger}}, \bibinfo {author} {\bibfnamefont {P.}~\bibnamefont {Ziemann}}, \bibinfo {author} {\bibfnamefont {E.}~\bibnamefont {Peiner}},\ and\ \bibinfo {author} {\bibfnamefont {A.}~\bibnamefont {Waag}},\ }\bibfield  {title} {\bibinfo {title} {Effect of annealing on the depth profile of hole concentration in \text{(Ga,Mn)As}},\ }\href {https://doi.org/10.1103/PhysRevB.71.205213} {\bibfield  {journal} {\bibinfo  {journal} {Phys. Rev. B}\ }\textbf {\bibinfo {volume} {71}},\ \bibinfo {pages} {205213} (\bibinfo
  {year} {2005})}\BibitemShut {NoStop}%
\bibitem [{\citenamefont {Moriya}\ and\ \citenamefont {Munekata}(2003)}]{14}%
  \BibitemOpen
  \bibfield  {author} {\bibinfo {author} {\bibfnamefont {R.}~\bibnamefont {Moriya}}\ and\ \bibinfo {author} {\bibfnamefont {H.}~\bibnamefont {Munekata}},\ }\bibfield  {title} {\bibinfo {title} {Relation among concentrations of incorporated \text{Mn} atoms, ionized \text{Mn} acceptors, and holes in \text{p-(Ga,Mn)As} epilayers},\ }\href {https://doi.org/10.1063/1.1559426} {\bibfield  {journal} {\bibinfo  {journal} {J. Appl. Phys.}\ }\textbf {\bibinfo {volume} {93}},\ \bibinfo {pages} {4603} (\bibinfo {year} {2003})}\BibitemShut {NoStop}%
\end{thebibliography}
    %

\end{document}